\documentclass{article}
\usepackage{graphicx}
\usepackage{epsf}
\usepackage{epsfig}
\usepackage{amssymb}
\usepackage{amsmath}
\usepackage{epstopdf}
\usepackage{amsfonts}
\usepackage{psfrag}

\newcommand{\heta}{\hat{\eta}}
\newcommand{\hmu}{\hat{\mu}}
\newcommand{\hxi}{\hat{\xi}}
\newcommand{\breta}{\bar{\eta}}
\newcommand{\brmu}{\bar{\mu}}
\newcommand{\brxi}{\bar{\xi}}

\DeclareMathOperator{\End}{End}
\DeclareMathOperator{\OSp}{OSp}
\DeclareMathOperator{\SO}{SO}
\DeclareMathOperator{\SU}{SU}
\DeclareMathOperator{\SSL}{SL}
\DeclareMathOperator{\OO}{O}

\DeclareMathOperator{\SL}{SL}

\DeclareMathOperator{\osp}{osp}
\DeclareMathOperator{\tr}{tr}

\DeclareMathOperator{\sdet}{sdet}
\DeclareMathOperator{\str}{str}
\DeclareMathOperator{\ssl}{sl}
\DeclareMathOperator{\ssp}{sp}

\DeclareMathOperator{\SP}{Sp}

\DeclareMathOperator{\sdim}{sdim}  
\DeclareMathOperator{\Sym}{Sym}
\DeclareMathOperator{\so}{so}

\makeatletter
\@addtoreset{equation}{section}
\makeatother

\def\heta{\hat{\eta}}
\def\bea{\begin{eqnarray}}
\def\eea{\end{eqnarray}}
\def\be{\begin{equation}}
\def\ee{\end{equation}}

\begin{document}

\def\wgta#1#2#3#4{\hbox{\rlap{\lower.35cm\hbox{$#1$}}
\hskip.2cm\rlap{\raise.25cm\hbox{$#2$}}
\rlap{\vrule width1.3cm height.4pt}
\hskip.55cm\rlap{\lower.6cm\hbox{\vrule width.4pt height1.2cm}}
\hskip.15cm
\rlap{\raise.25cm\hbox{$#3$}}\hskip.25cm\lower.35cm\hbox{$#4$}\hskip.6cm}}

\def\wgtb#1#2#3#4{\hbox{\rlap{\raise.25cm\hbox{$#2$}}
\hskip.2cm\rlap{\lower.35cm\hbox{$#1$}}
\rlap{\vrule width1.3cm height.4pt}
\hskip.55cm\rlap{\lower.6cm\hbox{\vrule width.4pt height1.2cm}}
\hskip.15cm
\rlap{\lower.35cm\hbox{$#4$}}\hskip.25cm\raise.25cm\hbox{$#3$}\hskip.6cm}}

\def\begeqar{\begin{eqnarray}}
\def\endeqar{\end{eqnarray}}

\begin{center}


\Large{A lattice approach to  the  conformal $\OSp(2S+2|2S)$   supercoset sigma 
model.}
\Large{Part II: The boundary spectrum.}

\vskip 1cm

Constantin Candu$^{(1)}$  and Hubert Saleur$^{(1,2)}$
\vspace{1.0em}


{\sl\small  Service de Physique Th\'eorique, CEA Saclay,\\
Gif Sur Yvette, 91191, France$^{(1)}$\\}
{\sl\small  Department of Physics and Astronomy,
University of Southern California\\
Los Angeles, CA 90089, USA$^{(2)}$\\}

\end{center}

\begin{abstract}
We consider  the partition function of the boundary 
$\OSp(2S+2|2S)$ coset sigma model on an annulus, 
based on the lattice regularization introduced in the companion paper. Using
results for  the action of $\OSp(2S+2|2S)$ and $B_L(2)$ on the
corresponding spin chain, as well as mini-superspace and small $g_\sigma^2$
calculations, we conjecture the full spectrum and set of degeneracies on the
entire critical line. Potential relationship with the $\OSp(2S+2|2S)$
Gross-Neveu model is also discussed. 
\end{abstract}

\section{Introduction}

This paper relies heavily on the results of its companion \cite{CanduSaleuri}. In the latter work, we have defined a lattice model which, we argued, is in the universality class of the $\OSp(2S+2|2S)$ coset sigma model, and carefully studied the decomposition of the Hilbert space of the corresponding quantum spin chain $V^{\otimes L}$ under the action of the supergroup $\OSp(2S+2|2S)$ and its non-semisimple  commutant $B_L(2)$. We shall now use these results to obtain information on the boundary spectrum of the model.

The paper is organized as follows. In section 2, various easily obtained results in the continuum limit are collected. Section 3 deals with the weak coupling limit in the bulk. Subsection 3.1 tackles the minisuperspace analysis, subsection 3.2 defines and studies the small coupling expansion around $g_\sigma^2=0$, subsection 3.3 discusses the exact structure of the theory as $g_\sigma^2\rightarrow 0$ and subsection 3.4 compares our perturbative approach with the one of Wegner formally extended to the $\OSp$ case. Section 4 exploits the algebraic structures of the lattice model elucidated in the first paper to formulate two essential conjectures about the boundary spectrum. Section 5 presents thorough  numerical checks of these conjectures. Section 6 discusses potential relationship with the $\OSp$ Gross Neveu models and our third conjecture. Section 7 contains our conclusions.

All notations are consistent with those in \cite{CanduSaleuri}. In addition, some of the new notations introduced in what follows are:
\begin{itemize}
\item $2\pi g=g_\sigma^2$, coupling constant of the sigma model
\item $\hbox{Vir}_B$ the chiral algebra obtained from the Brauer algebra in the
continuum limit
\end{itemize}

\section{Some immediate results for boundary partition functions in the continuum limit}

\subsection{The periodic partition function}
\label{sec:per_pf}

As discussed at the end of section 2 in \cite{CanduSaleuri}, the periodic partition function 
of the $\OSp(2S+2|2S)$ model  on the annulus is obtained by calculating the supertrace 
of the appropriate power of the transfer matrix, which translates 
geometrically by giving to all non 
contractible loops a weight two. This partition 
function is thus {\sl 
identical} with the one of the 6 vertex model. The continuum limit is 
well known to be described by a free compactified boson with 
Dirichlet
boundary conditions on either sides of the annulus. This bosonic 
degree of freedom comes from the interpretation of the 6 vertex model 
as a solid on solid model with height variables dual to the arrows. 
It is traditional to write the action for   this boson $\tilde{\varphi}$ as:
\begin{equation}
    S=
    {g\over 4\pi}\int d^{2}x(\partial_{\mu}\tilde{\varphi})^{2}
    \end{equation}
  with $\tilde{\varphi}$  quantized on a circle of circumference $2\pi$. 
Standard 
  results show the relation between  the coupling constant $g$ and 
the 
  anisotropy parameter of the lattice model:
 \begin{equation}
     \Delta=\cos\pi g
     \end{equation}
 so the case $\Delta=-1$ corresponds $g=1$ and $\Delta\rightarrow 1$ 
 to $g\rightarrow 0$.  The continuum limit of the partition function is then 
  \begin{equation}
      \tilde{Z}_{DD}={1\over \eta}\sum_{j} q^{gj^{2}}\label{Zdd}
      \end{equation}
where $j$ is integer for a lattice of even width, and $j$ is half an 
odd integer for a lattice of odd width. We have introduced the modular
parameter $q=e^{2\pi i \tau}$ to describe the annulus where $\tau = \frac{iT}{2L}$, $L$ is
the transverse length (so the hamiltonian or transfer matrix act on
the space $V^{\otimes L}$) and $1/T$ is the  length in
imaginary time (this $T$ must not  be confused with the transfer matrix, also denoted by the same letter in what follows). $\eta$ is the usual Dedekind function 
\begin{equation}
\eta(\tau)=q^{1/24}\prod_{n=1}^\infty (1-q^n).
\end{equation}

We note however that the loop partition function
for the $\OSp(2S+2|2S)$ model coincides with Neumann boundary 
conditions in terms of the orthosymplectic vector field (Dirichlet 
boundary conditions would turn into loops having open ends on the 
boundary). Therefore, the correspondence with the sigma model is best 
understood by turning to the dual of the field $\tilde{\varphi}$ whose 
action is 
\begin{equation}
    S=
    {1\over 4\pi g}\int d^{2}x(\partial_{\mu}\varphi)^{2}
    \end{equation}
 The partition function (\ref{Zdd}) can then be reinterpreted as the 
 partition function with Neumann boundary conditions
   \begin{equation}
       Z_{NN}={1\over \eta}\sum_{j} q^{gj^{2}}\label{Znn}
       \end{equation}

The identity of the partition function of the $\OSp(2S+2|2S)$ model 
and 
of the free boson in this case can of course be directly demonstrated 
at the level of the field theory. Indeed, the functional integral 
over the fundamental $\OSp(2S+2|2S)$ fields $\phi^{i}(\tau,\sigma)$
can be formally evaluated by using an integral representation for the constraint 
$\delta\left(J_{ij}\phi^{i}\phi^{j}-1\right)$. This leaves one with 
$2S+2$ identical bosonic, and $2S$ identical fermionic integrals, 
which cancel against each other, leaving two bosonic integrals. The 
constraint can then be reintegrated, leading to the action
\begin{equation}
    A={1\over 2g_{\sigma}^{2}}\int d^{2}x     
\left[(\partial_{\mu}\phi^{1})^{2}+(\partial_{\mu}\phi^{2})^{2}\right],     
~~~(\phi^{1})^{2}+(\phi^{2})^{2}=1\label{freebosonact}
    \end{equation}
From this we identify the Coulomb gas coupling
\begin{equation}
    2\pi g= g_{\sigma}^{2}
    \end{equation}
and $\phi^{1}=\cos\varphi,\, \phi^{2}=\sin\varphi$.

\subsection{The twisted partition function}
\label{sec:tw_pf}

In sec.~3.1 of \cite{CanduSaleuri} we have defined for the
spin chain $V^{\otimes L}$ a, so called,
quasiperiodic partition function depending on a supermatrix $D\in
\OSp(2S+2|2S)$ encoding the set of all possible boundary conditions.
Let us call \emph{twisted} the subset of boundary conditions such that
$\sdet D = -1$.
The simplest twisted boundary condition is of the form
\begin{equation} \label{twbc}
  \phi^1(\tau,\sigma+r)=\phi^2(\tau,\sigma),\quad
  \phi^2(\tau,\sigma+r)=\phi^1(\tau,\sigma), \quad
  \phi^i(\tau,\sigma+r)=\phi^i(\tau,\sigma), \; i\neq 1,2
\end{equation}
and the supermatrix $D$ encoding it is in fact the reflection $\rho$
described in \cite{CanduSaleuri}.

The twisted boundary condition in eq.~\eqref{twbc} defines a twisted
partition function $Z^\text{tw}$ which can be easily computed in the path integral
formalism. This,  again leads to the cancellation of $2S$ 
bosonic  integrals against the fermionic ones leaving the fields
$\phi^{1,2}$. After a simple 
rotation by $\pi/2$ in the space $\phi^{1,2}$ one can see that the
twisted boundary conditions in eq.~\eqref{twbc} are equivalent to antiperiodic  boundary conditions for 
the field  $\varphi$ itself.
The partition function for the compactified boson $\varphi$ is therefore:
\begin{equation*}
    Z^\text{tw}=[\hbox{det }(-\Delta_{DA})]^{-1/2}
    \end{equation*}
where $\Delta_{DA}$ is the Laplacian with Dirichlet boundary 
conditions in space and antiperiodic boundary conditions in (imaginary) time 
direction:
\begin{equation}\label{ptwpf}
    Z^\text{tw}=\sqrt{2\eta(\tau)\over
      \theta_{2}(\tau)}=\frac{\eta(\tau)}{\eta(2\tau)} =q^{-1/24}{1\over 
    \prod_{n=1}^{\infty}(1+q^{n})}=q^{-1/24}\prod_{n=1}^{\infty}(1-q^{2n-1}).
    \end{equation}
Note that this partition function is independent of $w$, ie of the
weight of the intersections.

The fermionic fields in eq.~\eqref{twbc} are periodic. One
can as well consider antiperiodic boundary conditions for fermions.
Therefore we shall distinguish between the periodic and antiperiodic
twisted partition functions, the one in eq.~\eqref{ptwpf} being periodic.

In the discrete loop model picture, imposing either periodic or antiperiodic twisted boundary conditions 
amounts to giving a weight 0 to the loops winding the annulus an
odd number of times. The difference resides in choosing 2 or $\tilde{N}:=4S+2$
for the loops winding the annulus an even (nonzero) number of times.

\subsection{The case $w=0$}
\label{sec:w=0}

In the case $w=0$, loops do not intersect and the partition function can be easily computed.
Noncontractible loops can wind only once around the system and get  a weight $\tilde{N}$ for (anti)periodic boundary conditions for bosons(fermions). The partition function can be written as
\begin{equation}\label{no_int_pf}
   Z=\sum_{j=0}^{\infty}D_{j}{q^{j^{2}}-q^{(j+1)^{2}}\over \eta(\tau)}
   \end{equation}
for an even lattice, while for an odd lattice, the sum runs $j$ half 
an odd integer (and thus 
adding odd and even widths gives the same sum for $j$ half integer). 
Here,
\begin{equation}
   D_{j}={\sinh(2j+1)\alpha\over\sinh\alpha}
   \end{equation}
and $\alpha$ is determined through  $D_{1/2}=\tilde{N}$ ($=4S+2$), the dimension of 
the fundamental. 
We then find the 
number of $(1,0)$ fields to be 
$D_{1}=1+2\cosh 2\alpha=\tilde{N}^{2}-1$, a result 
indicating the 
underlying $\SU(2S+2|2S)$ symmetry. Note that 
$D_{1}=D_{adj}+{\tilde{N}^2\over 2}$. The total number of $h=1$ fields 
in the open boundary partition function is thus the sum of the number of 
currents and ${\tilde{N}^{2}\over 
2}$ additional  primary fields in the rank 2 symmetric $\SU(2S+2|2S)$
tensor.

Note that the partition function for even lattice can also be written 
as
\begin{equation}
    Z=\sum_{j=0}^\infty (D_{j}-D_{j-1}){q^{j^{2}}\over \eta(\tau)}\label{wrongZ}
    \end{equation}
 with $D_{-1}:= 0$. Evaluating the difference of the two 
 dimensions leads to the simple formula:
 \begin{equation}
     Z=\frac{1}{\eta(\tau)}\theta_{3}\left({i\alpha\over\pi},2\tau\right).
     \end{equation}
 The modular transform immediately follows:
 \begin{equation}
     Z={1\over \sqrt{2}\eta\big(-\frac{1}{\tau}\big)}
     \sum_{n=-\infty}^{\infty} 
     \tilde{q}^{(n+i\alpha/\pi)^{2}/4}
     \end{equation}
where $\tilde{q}=e^{-2\pi i\frac{1}{\tau}}$ and exhibits complex exponents. These complex exponents can be traced back to the complex electric charges in the free boson theory (\ref{freebosonact}) necessary to give non contractible loops a weight greater than two. It is not 
clear of course  that the modular transform of the modified 
partition function should have a useful meaning in the sigma model CFT:
antiperiodic boundary conditions for fermions  in the space direction do not
have to be included for consistency of the model (unlike say, for the Majorana fermions of the Ising model).

There is yet another way of obtaining the partition function in eq.~\eqref{no_int_pf} at $w=0$ if one 
knows: i) the trace of the transfer matrix  restricted to an irrep of the Temperley Lieb algebra and ii) the 
decomposition of representations of $B_L(2)$ into irreps of the Temperley Lieb algebra.
According to sec.~4.1 of \cite{CanduSaleuri}, the partition function with most general boundary conditions (encoded in the matrix $D$) can be written in the form
\begin{equation*}
Z_D = \str_{V^{\otimes L}} T^\beta D^{\otimes L} = \sum_\lambda sc_\lambda(D)\chi'_\lambda(T^\beta),
\end{equation*}
where $\lambda\vdash L-2k,\, k=0,1,\dots,$ $sc_\lambda(D)$ are $\OSp(4|2)$ generalized symmetric functions and $\chi'_\lambda$ is the character of the standard $B_L(2)$ representation $\Delta_L(\lambda)$. On the other hand, according to the discussion in sec.~5.2 of \cite{CanduSaleuri}, the standard module $\Delta_L(\lambda),\lambda\vdash m=L-2k$  decomposes into a direct sum of irreps $D_L(m+2l)$
of the Temperley Lieb algebra as
\begin{equation*}
\Delta_L(\lambda) \simeq \bigoplus_{l=0}^k f_\lambda n(m,m+2l) D_L(m+2l)
\end{equation*}
where $f_\lambda$ is the number of standard Young tableaux of shape $\lambda$ and $n(m,l)$ are multiplicities.
If $c_j$ is the character of the Temperley Lieb irrep $D_L(2j)$ in the limit $L\to \infty$, then according to \cite{ReadSaleur07} (this is discussed in more details below)
\begin{equation*}
c_j\big(T^\beta(w=0)\big) = \frac{q^{j^2}- q^{(j+1)^2}} {\eta(\tau)}
\end{equation*}
and, therefore, in the continuum limit the partition function of the dense intersecting loop model at $w=0$ becomes
\begin{equation}\label{no_int_diff_pf}
Z(D) = \sum_{j\in \frac{\mathbb{N}}{2}}^\infty \left[\sum_{m=2j,2j-2,\dots}\left(\sum_{\lambda\vdash m }f_\lambda sc_\lambda(D) \right) n(m,2j)\right] \frac{q^{j^2}- q^{(j+1)^2 }} {\eta(\tau)}.
\end{equation}
In simple terms, huge degeneracies appear at this point. Indeed, since loop crossings are not allowed ($w=0$), different operators corresponding to different symmetries of the non contractible lines now become identical. 

 \section{Weak coupling results for the bulk theory}
 
 \subsection{Minisuperspace on the superphere $\OSp(2S+2|2S)/\OSp(2S+1|2S)$}

\label{minisup}

 The limit where $g_{\sigma}\rightarrow 0$  of the bulk spectrum can be analyzed using a 
 minisuperspace approximation. Such a strategy has proved extremely successful in recent analysis of WZW models on supergroups in particular \cite{Volkerreview,Volkerlast}.
 
  Indeed, consider  the sigma model on 
 a cylinder of circumference $r$ or, equivalently, at temperature 
 $T={1\over r}$. Doing a Wick rotation transforms  the space into a  basic circle $\sigma\equiv \sigma+r$, while the imaginary time runs 
 to infinity along the axis of the cylinder. At small $r$ - ie large 
 temperature - it is reasonable to neglect the fluctuations of the 
 fields in the transverse direction, and replace the fields 
 $\phi^i(\sigma,\tau)$ by $\phi^i(\tau)$. 

To be more precise, let us describe the problem in a hamiltonian
formalism. In general, one 
 has to deal with wave functions $\Psi$ which are functions of the 
 field configuration at a given time (or imaginary time), 
 $\Psi[\phi^i(\sigma)]$. In the minisuperspace limit, these become 
 functions of the $\sigma$ independent approximation of the fields, 
 ie {\sl functions on the target space itself}. If the sigma model 
of interest is a model on a (super)group, the wavefunctions become  
functions on that (super)group. 
The hamiltonian becomes a differential operator on these functions.

To see how this works and fix notations, consider briefly  the $\OO(2)$ 
action
\begin{equation}
  \label{eq:comp_bos}
   A={1\over 2g_{\sigma}^{2}}\int d^{2}x 
   \left[(\partial_{\mu}\phi^{1})^{2}+(\partial_{\mu}\phi^{2})^{2}\right]=
   {1\over 4\pi g}\int d^{2}x(\partial_{\mu}\varphi)^{2}
   \end{equation}
 with $\varphi$  the angle of the vector $\phi$, quantized on a 
circle of circumference $2\pi$. 
 The   minisuperspace approximation should be valid in the limit of 
 $g$ large. This corresponds to small 
 temperatures in the XY model, ie the limit where the free floating 
 vortex operators are strongly irrelevant.

 So, in the minisuperspace approximation, the action in eq.~\eqref{eq:comp_bos}
 \begin{equation*}
   A = \frac{1}{2 g_\sigma^2} \int d\,\tau \dot{\varphi}^2
 \end{equation*}
 yields the quantised hamiltonian
  \begin{equation*}
H={g_\sigma^2 T\over 2} \Pi^{2} = - \frac{g_\sigma^2 T}{2} \hat{\Delta}_1.
\end{equation*}
Here $\Pi$ is the canonical momentum associated with $\varphi$, the
equal time commutator is $[\Pi,\varphi] = 1/i$ and $\hat{\Delta}_1 =
d^2/d\varphi^2$ is the Laplacian on the circle.
  The hamiltonian $H$ has 
  eigenfunctions $\Psi_n(\phi)=e^{ni\varphi}$ with eigenenergies $E_{n}=g_{\sigma}^{2} T 
n^{2}/2$. Again, this approximation should become good when 
$g_{\sigma}^{2}$ is 
  large, so these dimensions are small and accumulate near the ground   state. 
  
  On the other hand, $H$ reads, in the Virasoro formalism
   \begin{equation*}
       H=2\pi T\left(L_{0}+\bar{L}_{0}-{c\over 12}\right).
       \end{equation*}
    The spectrum is thus, from the exact solution,
    \begin{equation}\label{o2sp}
	E=2\pi T\left( {e^{2}g_{\sigma}^{2}\over 4\pi}+{\pi\over 
	g_{\sigma}^{2}}m^{2}-{1\over 12}\right)
	    \end{equation}
   and coincides in the limit $g_{\sigma}^{2}$ small with the one obtained in 
the  minisuperspace limit indeed.
Note that the central charge being a contribution of order $O(1)$ to the
spectrum should not be visible in the minisuperspace approximation.

Let us now apply these ideas to the simplest non trivial model of 
our  series, namely the supersphere $S^{3|2}$.
We shall not dwell here on the subtleties related to rigorous definition of
the supersphere as a supermanifold in the sense of mathematicians.
Instead, we prefer to define it directly as the coset space $S^{3|2} :=
\OSp(4|2)/\OSp(3|2)$.

To be more specific let us fix some notations.
Let $B_L$ be some Grassman algebra with a large enough number of
generators $L$.
Consider the linear space $\mathbb{C}^{4|2}$ over $B$ composed of points $X$,
which can be parametrized by four even coordinates $X^i=x^i\in B$, $i=0,1,2,3$
and two odd coordinates $X^\alpha =\eta^\alpha\in B$, $\alpha=1',2'$.

$\mathbb{C}^{4|2}$ becomes a supereuclidian linear space $E^{4|2}$ if
we endow it with a scalar product which is defined as follows:
for two points $X,Y\in \mathbb{C}^{4|2}$ with coordinates
$x^i,\eta^\alpha$ and, respectively, $y^i,\xi^\alpha$ put 
\begin{equation*}
\label{eq:sc_prod}
X\cdot Y = X^p J_{pq}Y^q = x^i J_{ij} y^j + \eta^\alpha J_{\alpha \beta}\xi^\beta =
\sum_{i=0}^3 x^i y^i + \eta_1 \xi_2 - \eta_2 \xi_1,
\end{equation*}
where $p,q = 0,1,2,3,1',2'$.
To distinguish between even and odd components we introduce the grading
function $|\cdot |$ which is zero evaluated on even indices and one on
the odd ones, e.g. $|i|=0, |\alpha|=1$.

It is then natural to associate to each $X\in E^{4|2}$ a
point $X^*$ in the dual space  by the usual index
lowering procedure $X_p = X^q J_{qp}$  in such a way that the scalar
product $X\cdot Y $ becomes $X^*(Y) = X_p Y^p$ and to each endomorphism $M$ of
$E^{4|2}$ the dual(transpose) $M^*$ by the correspondence $Y = M
X\Rightarrow Y^* = X^* M^*$. In matrix components $(M^*)_q^{~p}=M^p_{~q}(-1)^{|q||p|}$.

An element of $\OSp(4|2)$ is an element of $\End E^{4|2}$
orthogonal with respect to the scalar product in
eq.~\eqref{eq:sc_prod}, that is in matrix notations
\begin{equation}
  \label{eq:osp42_def}
  M^* M = I.
\end{equation}
Note that, so defined, the supergroup $\OSp(4|2)$ is noncompact because it
contains as a subgroup the group $\SP(2)\simeq \SL(2)$ of
transformation of $\eta^\alpha$
only.

The supergroup $\OSp(3|2)$ is then realized as the subgroup
of $\OSp(4|2)$ stabilizing the line $x^0$.
Therefore, a point on $S^{3|2}$ has coordinates of the form
$X^p = M^p_{~0}$ for some $M\in \OSp(4|2)$.
One can see from eq.~\eqref{eq:osp42_def} that the coordinates
of the points $X\in S^{3|2}$ satisfy the equation
\begin{equation}
\label{eq:s32_def}
X_pX^p = \sum_{i=0}^3(x^i)^2 + 2\eta^1\eta^2 = 1,
\end{equation}
giving the embedding of $S^{3|2}$ into $E^{4|2}$.
The solutions of eq.~\eqref{eq:s32_def} can be parametrized as follows
   \begin{align}
\label{eq:resc}
       &x^{i}=n^{i}(1-\eta^{1}\eta^{2})\\ \label{eq:s3}
       &\sum_{i=0}^{3}(n^{i})^{2}=1.
       \end{align}
Note that one can introduce the spherical or Euler angles to parametrize the
$n^i$'s in the same way as  for
the embedding of $S^3$ into $\mathbb{R}^3$.
However, when the body of some component $n^i$ vanishes, that is
$b(n^i)=0$, only the square of the soul of $n^i$
is fixed by eq.~\eqref{eq:s3}, which is not enough to uniquely
determine the soul itself. Therefore, the parametrisation with
spherical or Euler angles does not give the full set of solutions when
$b(n^i)=0$ for some $i$.

The infinitesimal distance element is obviously
\begin{equation*} 
  dX_pdX^p = 2(1-\eta^1\eta^2)d\eta_1d\eta_2+ 
  (1-2\eta_1\eta_2)\sum_{i=0}^3 (dn^i)^2.
\end{equation*}  
Solving the constraints for $n^i$, one can extract the metric tensor $g_{ab}$
on $S^{3|2}$.

Then, in terms of fields $n^i, \eta^\alpha$, the $S^{3|2}$ sigma model
field theory action
   \begin{equation}
\label{act}
       A =\frac{1}{2 g_\sigma^2} \int d^2\, x \left( \sum_{i=0}^{3}(\partial_{\mu}x^i)^{2}+
       2\partial_{\mu}\eta^1\partial_{\mu}\eta^2 \right)
       \end{equation}
    becomes
   \begin{equation}
       A={1\over 2g_{\sigma}^{2}}\int d^2\,x 
       \left( 2(1-\eta^{1}\eta^{2})
       \partial_{\mu}\eta^{1}\partial_{\mu}\eta^{2}+(1-2\eta^{1}\eta^{2})
       \sum_{i=0}^{3}(\partial_{\mu}n^{i})^{2}\right).
       \label{normact}
       \end{equation}

The square root of $g= \sdet g_{ab}$ fixes the invariant measure on
the supersphere $dS^{3|2}=[(1-2\eta_1\eta_2)d\eta_1 d\eta_2] dS^3$ in the
path integral formalism, where $dS^3$ denotes the
invariant measure on the target space $S^3$ for the fields $n^i$.


We shall use the isomorphism $f:S^3\rightarrow SU(2)$ to
give a parametrization of (almost) all the sphere $S^3$.
Thus, if $n^i$ are the coordinates of a point on $S^3$, then the
corresponding element of $SU(2)$ is
$G= f(n^0,n^1,n^2,n^3)= n^0 + \sum_{a=1}^3\sigma^a n^a$,
where $\sigma^a$ are Pauli matrices.

Next, recall that there is an isomorphism  $g: \SO(4) \rightarrow SU(2)\otimes
SU(2)/\mathbb{Z}_2$.  Thus, if $g(R) = G_L\otimes G_R$ then the action of
$R$ on $n^i$'s can be represented as $f(R n) = G_L f(n) G^{\dag}_R$, where on
the right hand side we have a matrix multiplication.

In what follows we shall use the Hopf parametrisation of $SU(2)$,
which shall prove more comfortable then the usual parametrisation with
Euler angles
\begin{equation}
\label{hp}
  G=\begin{pmatrix}
    e^{i \xi_1}\cos\mu & e^{i\xi_2} \sin\mu\\
    -e^{-i \xi_2} \sin\mu& e^{-i \xi_1}\cos\mu
\end{pmatrix},
\end{equation}
with $0\leq b(\mu) \leq \pi/2$ and $0\leq b(\xi_1),b(\xi_2)<2\pi$,
where $b:B_L \rightarrow \mathbb{R}$ denotes the body map.
According to the remark made before, the points
$b(\mu),b(\xi_1),b(\xi_2)=\pi \mathbb{Z}/2$ are singular for the
parametrization in eq.~\eqref{hp}.

Using the Hopf parametrisation of $S^3$ we get 
\begin{equation*}
   \sum_{i=0}^{3}(\partial_{\mu}n^{i})^{2} =
   \frac{1}{2}\tr \left( \partial G^\dagger \partial G
   \right)=
(\partial \mu)^{2} + \cos^2\mu
     (\partial \xi_1)^{2} + \sin^2\mu (\partial \xi_2)^{2}.
\end{equation*}

The classical minisuperspace hamiltonian provided by
eq.~\eqref{normact} is then
\begin{equation}
\label{cl_ham}
H =
\frac{g_\sigma^2 T}{2} \left[ 2(1 +
  \eta^1\eta^2) \Pi_{\eta^1} \Pi_{\eta^2}+
  (1+2\eta^1\eta^2)\left(\Pi_\mu^2+ \frac{1}{\cos^2\mu}\Pi_{\xi_1}^2 +
    \frac{1}{\sin^2\mu}\Pi_{\xi_2}^2 \right)\right].
\end{equation}
 
To quantize this hamiltonian one has to write the evolution operator
in the path integral formalism and then derive the Schrodinger equation it
satisfies by propagating the wave function for an infinitesimal amount
of time. A shortcut to the correct final result is the
ordering prescription for coordinate and canonical momenta
yielding an invariant second order differential operator, that
is the Laplace operator on $S^{3|2}$
   \begin{equation}
       g^{ba}\Pi_{a}\Pi_{b}\rightarrow \frac{1}{\sqrt{g}} 
       \hat{\Pi}_{a}g^{ba}\sqrt{g}\hat{\Pi}_{b}.
       \label{order}
       \end{equation}
For the parametrisation in eq.~\eqref{hp} the
nonvanishing components of the metric tensor $g_{ab}$ are
\begin{equation}
\label{met}
g_{\mu\mu}=1-2\eta^1\eta^2,\quad g_{\xi_{1}\xi_{1}}=
(1-2\eta^1\eta^2)\cos^2\mu, \quad g_{\xi_2\xi_2}=(1-2\eta^1\eta^2)\sin^2\mu,
\quad g_{\eta^1\eta^2} = (1-\eta^1\eta^2).
\end{equation}
According to eqs.~\eqref{order} and \eqref{met} the quantized
hamiltonian becomes~\footnote{As it is typical for quantum mechanics, this hamiltonian is defined up to an
  arbitrary constant. This constant has its origin in the
  arbitrariness of the measure of the
  regularized path integral.}
 \begin{equation}
\label{eq:quant_ham}
         \hat{H}= - \frac{T g_\sigma^2}{2}\hat{\Delta}_{3|2}= T g_{\sigma}^{2}
       \left[-(1+\eta^1\eta^2)
       \partial_{\eta^{1}}\partial_{\eta^{2}}+{1\over
2}\eta^{1}\partial_{\eta^{1}}
       +{1\over
       2}\eta^{2}\partial_{\eta^{2}}+2(1+2\eta^{1}\eta^{2})\Delta_{ SU(2)}\right],
       \end{equation} 
where $\Delta_{SU(2)}$ is the Laplace operator on $SU(2)$ normalized
to have eigenvalues $j(j+1)$ with $j$ integer or half integer
 \begin{equation}
     \Delta_{SU(2)}:=
\frac{\partial^2}{\partial \mu^2} + 2 \cot2\mu
\frac{\partial}{\partial \mu} + \frac{1}{\cos^2
  \mu}\frac{\partial^2}{\partial\xi_1^2}+ \frac{1}{\sin^2\mu}\frac{\partial^2}{\partial\xi_2^2}.
     \end{equation}
Note that one can get the same quantum hamiltonian starting
from the Laplace operator in $E^{4|2}$
\begin{equation}
  \label{elap}
  \Delta_{4|2}:=J^{qp}\frac{\partial}{\partial X^p}\frac{\partial}{\partial X^q}
\end{equation}
by first making the change of coordinates
\begin{equation*}
  X^i = R x^i, \quad X^\alpha = R \eta^\alpha,
\end{equation*}
where $R = \sqrt{X_p X^p}$ and $x^i,\eta^\alpha$ are as in
eq.~(\ref{eq:resc},\ref{eq:s3}), and then subtracting the radial part
\begin{equation*}
  \hat{\Delta}_{3|2} = R^2\left(\Delta_{4|2}- \frac{\partial^2}{\partial
    R^2} -  \frac{\partial}{\partial R}\right).
\end{equation*}
The operator  $\hat{\Delta}_{3|2}$ is also the second order Casimir 
in the regular representation of $\OSp(4|2)$ in the space of functions
on $S^{3|2}$.
 
It is easy to diagonalize $\hat{H}$ by hand. One finds the 
 eigenvalues 
 \begin{equation}
\label{eig}
     E_{n}=2 T g_{\sigma}^{2} j^{2}={ Tag_{\sigma}^{2}\over 2}n^{2}
     \end{equation}
 where $j$ is the $SU(2)$ spin, $n=2j$ is the $SU(2)$ Dynkin label.

The  eigenfunctions are of four types:
 \begin{eqnarray}
     (1-n\eta^{1}\eta^{2})f_{mm'}^{j};~~~
     (1+n\eta^{1}\eta^{2})f_{mm'}^{j-1};~~~
     \eta^{1}f_{mm'}^{j-1/2};~~~
     \eta^{2}f_{mm'}^{j-1/2}	\label{egfunc}
     \end{eqnarray}
where the $f_{mm'}^{j}$ are the $(2j+1)^{2}$ eigenfunctions of 
$\Delta_{SU(2)}$ with eigenvalue $j(j+1)$. They form a basis of the 
left or right regular representation of $SU(2)$. 
   
The dimension of the eigenvalue space is, for $j\geq 1/2$, 
thus made of 
 $[2(j-1)+1]^{2}+(2j+1)^{2}$ bosons and $2\times 
 [2(j-1/2)+1]^{2}$ fermions, leading to a superdimension 2 
 independent of $j$, and a dimension of $4 n^{2}+2, \, j\geq 1$.
In the $j=0$ case there is only one eigenfunction which is a constant,
while in the $j=1/2$ case there are two fermionic eigenfunctions
$\eta^1,\eta^2$ and four bosonic eigenfunctions
$(1-\eta^1\eta^2)G^{ij}$, where $G^{ij}$ are the matrix elements of
the $SU(2)$ matrix in eq.~\eqref{hp}.


The conformal weights in the minisuperspace approximation
can be computed by identifying the quantum evolution operator per unit
of time $e^{-\hat{H}}$ for a
particle moving on $S^{3|2}$ with the transfer matrix
$q^{L_0-\frac{1}{24}}\bar{q}^{\bar{L}_0-\frac{1}{24}}$ of the field
theory sigma model. 
Form eq.~\eqref{eig} and the identification $E_n = 4\pi T h_n$ the
conformal weights in the minisuperspace approximation will be 
\begin{equation*}
  h_n = \frac{g_\sigma^2 n^2}{8 \pi} = \frac{g n^2}{4}.\label{sym}
\end{equation*}
These are exactly the XY conformal weights in eq.~\eqref{o2sp} in the limit $g_\sigma \to 0$.

Let us point in the end how a similar minisuperspace analysis can be carried out in
the case of all supergroups $\OSp(2S+2|2S)$.
The supereuclidian space $E^{2S+2|2S}$ is defined as follows.
For the sake of notation, a point in  $\mathbb{C}^{2S+2|2S}$ is parametrized by
the set of even coordinates $X^i=x^i,\,i=1,\dots,2S+2$ and odd
coordinates $X^\alpha = \eta^\alpha$, $X^{S'+\alpha} = \bar{\eta}^\alpha$,
$\alpha = 1',\dots,S'$.
The scalar product between two points $X,Y\in E^{2S+2|2S}$ with
coordinates $x^i,\eta^\alpha,\bar{\eta}^\alpha$ and, respectively,
$y^i,\xi^\alpha,\bar{\xi}^\alpha$ is set to 
\begin{equation*}
  X\cdot Y = X^p J_{pq}Y^q = x^i J_{ij}y^i + \eta^\alpha
  J_{\alpha\beta}\xi^\beta = \sum_{i=1}^{2S+2}x^i y^i + \sum_{\alpha =
  1}^S \big(\bar{\eta}^\alpha \xi^\alpha - \eta^\alpha\bar{\xi}^\alpha\big).
\end{equation*}
The Laplacian in $E^{2S+2|2S}$ is defined as in eq.~\eqref{elap} and
the remarks made above remain valid for the Laplacian
$\hat{\Delta}_{2S+1|2S}$ on the superphere $S^{2S+1|2S}$.
 
The quantised hamiltonian will be $\hat{H} = -\frac{T
  g_\sigma^2}{2}\hat{\Delta}_{2S+1|2S}$ and 
\begin{equation}
  \label{genlap}
  \hat{\Delta}_{2S+1|2S}= \frac{1}{1- \eta^2}\hat{\Delta}_{2S+1|0}+ \Delta_{0|2S}
  - D_\eta^2,
\end{equation}
where   $\eta^2 := J_{\alpha\beta}\eta^\alpha\eta^\beta$,
$\Delta_{0|2S} :=
J^{\alpha\beta}\partial_{
  \eta^\beta}\partial_{\eta^\alpha}$ and $D_\eta :=
\eta^\alpha \partial_{\eta^\alpha}$.

We shall search for eigenfunction of the hamiltonian in the functional
space $L_2(S^{2S+12S}):= L_2(S^{2S+1})\otimes
\bigwedge (\eta)$, where $\bigwedge(\eta)$ is the Grassman algebra in
the generators $\eta^\alpha,\bar{\eta}^\alpha$.

Let $\mathsf{S}(x)$ denote the polynomial algebra in the
variables $x^i$ and consider the natural filtration of
$\mathsf{S}(x)$, seen as a vector space, by the homogeneous degree of
its elements
\begin{equation*}
  \mathsf{S}(x) \simeq \bigoplus_{n\in \mathbb{N}}  \mathsf{S}^n(x).
\end{equation*}
Counting all monomials of homogeneous degree $n$ is not hard to see
that $\dim \mathsf{S}^n(x) = C_{2S+1+n}^n$. Clearly the vector space
$\mathsf{S}^n(x)$ provides a $\SO(2S+2)$ representation which is
equivalent to a totally symmetric tensor of
rank $n$.
Let $\mathfrak{H}^n(x)\subset \mathsf{S}^n(x)$ denote the vector subspace of harmonic
polynomials. We shall need the following well known facts
\begin{align}\label{even_harm}
&\mathfrak{H}^n(x)\simeq \mathsf{S}^n(x) \big/ x^2
\mathsf{S}^{n-2}(x)\\ \notag
&h_0(n):=\dim \mathfrak{H}^n_{2S+2|0} = C_{2S+1+n}^{n}-C_{2S-1+n}^{n-2}\\\label{even_dec}
&L_2(S^{2S+1})\simeq \bigoplus_{n\in  \mathbb{N}}\mathfrak{H}^n(x)\\  \notag
&\hat{\Delta}_{2S+1|0} b_n(x) = -n(n+2S)b_n(x),\quad b_n(x)\in
\mathfrak{H}^n(x),\quad X^2 = 1,
\end{align} 
see \cite{Vilenkin}. Note that $\hat{\Delta}_{2S+2|0}$ has the same
eigenvalues as the $\SO(2S+2)$ second order Casimir evaluated on a
traceless symmetric tensor or rank $n$, while $h_0(n)$ is the number
of its independent components. Therefore the $\SO(2S+2)$
representation provided by the vector space $\mathfrak{H}^n(x)$ is
equivalent to a traceless symmetric tensor of rank $n$.

In order to generalize these results for the odd case consider the decomposition
\begin{equation}\label{grad_odd}
  \bigwedge(\eta) \simeq \bigoplus_{m=0}^{2S}\protect{\bigwedge}^m(\eta).
\end{equation}
It is most useful to exploit the fact that
$\protect{\bigwedge}^m(\eta)$ provide a $\SP(2S)$ representation
equivalent to a totally antisymmetric tensor of rank $m$. It is well
known that the representations provided by the action of $\SP(2S)$ on
antisymmetric tensors of rank $m$ and $2S-m$ are equivalent for
$m=0,\dots, S$, see
\cite{hamermesh}. This observation is at the origin of the following isomorphism of
vector spaces~\footnote{This is more then an isomorphism of vector spaces,
  it is actually an isomorphism of $\SP(2S)$ modules.}
\begin{equation}\label{shortcut}
  \protect{\bigwedge}^{2S-m}(\eta)\simeq \eta^{2(S-m)}\protect{\bigwedge}^m(\eta).
\end{equation}
In particular this means that
\begin{equation}\label{fundeq}
\dim \eta^{2 p} \protect{\bigwedge}^m(\eta) = \dim
\protect{\bigwedge}^m(\eta) = C^m_{2S},\quad p=1,\dots, S-m.
\end{equation}
Let $\mathfrak{H}^m(\eta)$ denote the vector space of harmonic
polynomials (with respect to the Laplacian $\Delta_{0|2S}$) of
homogeneous degree $m\leq S$. Then, using $\dim \eta^2
\protect{\bigwedge}^{m-2}(\eta) = \dim
\protect{\bigwedge}^{m-2}(\eta),\, m\leq S$ one can prove exactly as
in the case of eq.~\eqref{even_harm} the isomorphism
\begin{align*}
  &\mathfrak{H}^m(\eta)\simeq \protect{\bigwedge}^m(\eta)\bigg/ \eta^2
  \protect{\bigwedge}^{m-2}(\eta)\\
&h_1(m):= \dim \mathfrak{H}^m(\eta)= C^m_{2S}-C^{m-2}_{2S}.
\end{align*}
Note that $h_1(m)$ is the number of components of a traceless
antisymmetric $\SP(2S)$ tensor of rank $m$. Therefore
$\mathfrak{H}^m(\eta)$ is an irreducible $\SP(2S)$ representation.
The decomposition of $\SP(2S)$ antisymmetric  tensors
into irreps is reflected by the following relation derived from eq.~\eqref{fundeq}
\begin{equation}\label{interm}
  \protect{\bigwedge}^m(\eta) \simeq \mathfrak{H}^m(\eta)\oplus \eta^2
  \mathfrak{H}^{m-2}(\eta)\oplus \dots,\quad m\leq S
\end{equation}
Using eqs.~(\ref{grad_odd},\ref{fundeq},\ref{interm}) one finally arrives at
the analog of eq.~\eqref{even_dec}
\begin{equation}\label{odd_decomp}
\bigwedge (\eta) \simeq \bigoplus_{m=0}^S\bigg( \mathfrak{H}^m(\eta)
\oplus \eta^2  \mathfrak{H}^m(\eta) \oplus 
\dots \oplus \eta^{2(S-m)} \mathfrak{H}^m(\eta)\bigg) \simeq
\sum_{m=0}^S \mathbb{C}[\eta^2]\big/ \eta^{2(S-m)} \otimes \mathfrak{H}^m(\eta),
\end{equation}
where $\mathbb{C}[t]$ is the polynomial algebra over $\mathbb{C}$ in
one indeterminate $t$.
The $\SP(2S)$ second order Casimir in $\bigwedge(\eta)$ regular
representation is $\hat{\Delta}_{0|2S}:=\eta^2 \Delta_{0|2S}+2(S+1)D_\eta -
D^2_{\eta}$. One can check that all $\eta^{2p} \mathfrak{H}^m(\eta)$
belong to the same eigenspace of $\hat{\Delta}_{0|2S}$ corresponding to the
eigenvalue $-m(m-2S-2)$ and that indeed $\sum_{m=0}^S
(S-m+1)\big(C^m_{2S}-C^{m-2}_{2S}\big)=2^{2S}$.

Now we are ready to search for the whole set of eigenfunction of
$\hat{H}$.
From the remark made above, the Laplacian $\hat{\Delta}_{2S+1|2S}$
is the second order Casimir in the left (or right) regular
representation in $L_2(S^{2S+1|2S})$.
Therefore, it commutes with the action of $\OSp(2S+2|2S)$. This means
that, for an eigenvalue $\lambda$ of  $\hat{\Delta}_{2S+1|2S}$, the
eigenvalue subspace $L_\lambda\in L_2(S^{2S+1|2S})$ is
 $\OSp(2S+2|2S)$ invariant and, consequently, admits a $\SO(2S+2)\times
 \SP(2S)$ invariant decomposition of
 the form~\footnote{Eqs.~(\ref{even_dec},\ref{odd_decomp}) are, in
   fact, module isomorphisms.}
\begin{equation}\label{no_mult}
  L_\lambda \simeq \bigoplus_{p,m,n} d_{mn}\; \mathbb{C}[\eta^2]/\eta^{2(S-m+1)}\otimes 
  \mathfrak{H}^m(\eta)\otimes \mathfrak{H}^n(x),
\end{equation}
where $d_{mn}$ are multiplicities.
This being said, the most appropriate ansatz for the eigenfunction of
$\hat{\Delta}_{2S+1|2S}$  is 
\begin{equation}\label{ansatz}
  F(X) = g(\eta^2)f_m(\eta)b_n(x),\quad f_m(\eta)\in
  \mathfrak{H}^m(\eta),\; b_n(x)\in \mathfrak{H}^n(x),\; X^2=1,
\end{equation}
where $g(\eta^2) = \sum_{k=0}^{S-m}g_k \eta^{2k}$.
Plugging the ansatz \eqref{ansatz} into $\hat{\Delta}_{2S+1|2S}F(X)=
\lambda F(x)$ one gets a recurrence relation
\begin{equation*}
-4(k+1)(S-m-k) g_{k+1} = [(2k+m)^2+\lambda ] g_k+ n(n+2S)\sum_{l=0}^k g_l
\end{equation*}
which ends with a polynomial equation of degree $S-m+1$ in $\lambda$
\begin{equation*}
  [(2S-m)^2+\lambda] g_{S-m}+n(n+2S) \sum_{k=0}^{S-m}g_k = 0
\end{equation*}
with the solutions
\begin{equation}\label{sol_eg}
  \lambda^p_{mn} = -(m+n+2p)^2,\quad p=0,\dots,S-m+1.
\end{equation}
If $p\neq p'$ then   $\lambda^p_{mn}\neq \lambda^{p'}_{mn}$.
This means that there are no multiplicities in eq.~\eqref{no_mult}, that is all $d_{mn}=1$.
We have not succeeded to arrive at a compact analytical form for the
coefficients $g_k$. However, by replacing $\eta^2$ with a complex
indeterminate $t$, it is not hard see that the eigenvalue problem for
$g(\eta^2)$ is closely related to a, somewhat more familiar, eigenvalue problem
\begin{equation*}
  4t(1-t)h''(t) - 4\big[(m+1)t +S-m\big]h'(t) -\bigg[m^2 +
  \frac{n(n+2S)}{1-t}\bigg]h(t)=\lambda h(t),
\end{equation*}
which has the same solution for eigenvalues and $h_k=g_k,\,
k=0,\dots,S-m$ if $h(t)= \sum_{k=0}^\infty h_k t^k$.

It is interesting to see how the eigenfunctions $F^p_{mn}(X) =
g^p_{mn}(\eta^2)f_m(\eta) b_n(x),\,X^2=1$ corresponding to the same
eigenvalue $\lambda=\lambda^p_{mn}$ organize into a  $\OSp(2S+2|2S)$
representation. Suppose first that $\lambda = -(S+n)^2$. Then
the decomposition \eqref{no_mult} can be nicely represented in the
form bellow
\begin{center}
  \begin{tabular}[h]{cccccccc}
    & & & $F^0_{S,n}$ & & & & \\
    & & $F^0_{S-1,n+1}$ & & $F^1_{S-1,n-1}$ & & &\\
    & $F^0_{S-2,n+2}$ & & $F^1_{S-2,n}$  & & $F^2_{S-2,n-2}$ & & \\
    $F^0_{S-3,n+3}$ & & $F^1_{S-3,n+1}$ & & $F^2_{S-3,n-1}$ & &  $F^3_{S-3,n-3}$ &\\ 
    & $\cdots$ & &  $\cdots$ & & $\cdots$ & &  
  \end{tabular}
\end{center}
The eigenfunctions in the same row(column) have the same
$\SP(2S)$ ($\SO(2S+2)$) highest weight, which decreases from top to
bottom(left to right).

The eigenfunction $F^0_{Sn}$ at the top has the unique $\OSp(2S+2|2S)$
highest weight in $L_\lambda$.
In the notations of our previous paper \cite{CanduSaleuri} this
highest weight is $\Lambda = \delta_1+\dots+\delta_S+n \epsilon_1$ and   
is represented by a one row Young tableaux of width $S+n$.
The explicit form of the highest weight vector of $L_\lambda$ is
$F^0_{Sn}(X) = \eta^1\dots\eta^S H[(x^1)^n]$, where $H:
\mathsf{S}^n(x)\rightarrow \mathfrak{H}^n(x)$ is the canonical
projection map~\footnote{We have addopted the convention that $\eta^\alpha(\bar{\eta}^\alpha)$ have weight
  $\delta_\alpha(-\delta_\alpha)$ and $x^1$ has the highest $\SO(2S+2)$
  weight.}. Even more explicitly $H[(x^1)^n] = |x|^n C^S_n(x^1/|x|)$,
where $C^S_n$ are Gegenbauer polynomials.
Moreover, the value of the second order Casimir in the $\OSp(2S+2|2S)$
irreps $g(\Lambda)$ with highest weight $\Lambda$ is also $-(n+S)^2$.
Given that all other $\OSp(2S+2|2S)$ irreps with the same value
$-(n+S)^2$ of the Casimir  have higher highest weights,
we have proved that $L_\lambda$ is
equivalent to $g(\Lambda)$.

When $\lambda = -m^2,\, m\leq S$ the structure of $L_\lambda$ is the
same as in the tableau above, except that the highest weight vector is
$F^0_{m0} = \eta^1\cdots\eta^m$.

The dimension of $L_\lambda$ can be computed as follows
\begin{align*}
  \dim L_\lambda &= \sum_{p=0}^S \sum_q h_1(S-p-2q)[h_0(n+p)+h_0(n-p)]\\
  &= \sum_{p=0}^{2S}C^p_{2S}h_0(l-p)=
     \sum_{p=0}^{2S}C^p_{2S}\big[C^{l-p}_{2S+1+l-p}-C^{l-p-2}_{2S-1+l-p}\big]\\
&= \oint \frac{dz}{2\pi
  i}\sum_{p=0}^{2S}\sum_{q=0}^\infty C^p_{2S}z^p  C^q_{2S+1+q} z^q
\left(\frac{1}{z^{l+1}}- \frac{1}{z^{l-1}}\right)\\
&=  \oint\frac{dz}{2 \pi i}\frac{(1+z)^{2S}}{(1-z)^{2S+2}}
\left(\frac{1}{z^{l+1}}-\frac{1}{z^{l-1}}\right)= sc_l(\tilde{I}),
\end{align*}
where we have supposed again that $\lambda = -l^2,\, l=n+S$ and
$\tilde{I}$ is an even supermatrix which is identity in the
$\SO(2S+2)$ sector and minus identity in the $\SP(2S)$ sector.
So, according to the discussion in our previous paper
\cite{CanduSaleuri} on the
generalized $\OSp(2S+2|2S)$ Schur functions $sc_\mu$,
 we see that $\dim L_\lambda$ is equal to the number of components of
 a supersymmetric tensor of rank $l$.

The superdimension of $L_\lambda$ can be computed in essentially the
same way
\begin{align*}
  \sdim L_\lambda &=   \sum_{p=0}^S \sum_q (-1)^{S-p-2q}
  h_1(S-p-2q)[h_0(n+p)+h_0(n-p)] = \sum_{p=0}^{2S}(-1)^l C^p_{2S}h_0(l-p)\\
&= \oint\frac{dz}{2 \pi
    i}\frac{(1-z)^{2S}}{(1-z)^{2S+2}} 
\left(\frac{1}{z^{l+1}}-\frac{1}{z^{l-1}}\right)= 2.
\end{align*}

Let us end this section with an illustration of how the $\OSp(4|2)$ supersymmetric
tensors break down into fields of the form \eqref{egfunc}. This is
obvious in the case of the tensor of rank 1.
The components of the rank two traceless supersymmetric tensor are
\begin{equation*}
  \Sigma^{pq} = X^p X^q - \frac{1}{2}J^{pq}.
\end{equation*}
The eight  fields $\eta^\alpha f^{1/2}_{m,m'}$ in eq.~\eqref{egfunc} correspond to the
fermionic components $\Sigma^{\alpha i} = \eta^\alpha n^i$.
The bosonic components $\Sigma^{ij}$ can also be written in a manifest
$\SO(4)\times\SP(2)$ invariant form
\begin{equation*}
 \Sigma^{ij} = (1-2\eta^1\eta^2)n^in^j -\frac{1}{2}\delta^{ij} =
 (1-2\eta^1\eta^2)\left(n^in^j - \frac{1}{4}\delta^{ij}\right) -
 \frac{1}{4}(1+2\eta^1\eta^2)\delta^{ij}.
\end{equation*}
The remaining component
$\Sigma^{\alpha\beta}=J^{\beta\alpha}(1+2\eta^1\eta^2)/2$ is not independent
because of the vanishing trace condition.
The same argument can be repeated in the case of tensors of higher
rank. For instance the traceless supersymmetric tensor of rank 3 is of
the form
\begin{equation*}
  \Sigma^{pqr}=X^pX^qX^r - \frac{1}{4}(X^pJ^{qr}+ (-1)^{|p||q|}X^q
  J^{pr}+ X^r J^{pq}).
\end{equation*}
After some combinatorics one can prove that the nonzero
components of the supersymmetric tensor of rank $l=2j$ can be written in
an $\SO(4)\times \SP(2)$ manifestly invariant way as follows
\begin{align*}
  &\Sigma^{i_1\dots i_{l-2}i_{l-1}i_l} = (1-l\eta^1\eta^2)S^{i_1\dots
    i_{l-2}i_{l-1}i_l}-\frac{1+l\eta^1\eta^2}{2l(l-1)}\left(S^{i_1\dots
    i_{l-2}}\delta^{i_{l-1}i_l}+ \dots\right)\\
 &\Sigma^{i_1\dots i_{l-1}\alpha} =S^{i_1\dots
   i_{l-1}}\eta^\alpha,\quad
 \Sigma^{i_1\dots i_{l-2}\alpha\beta} =
 -\frac{1+l\eta^1\eta^2}{2(l-1)}S^{i_1\dots i_{l-2}}J^{\alpha\beta}
\end{align*}
where $S^{i_1,\dots,i_l}$ are the traceless $\SO(4)$ tensors of rank
$l$, that is the tensor form of the functions $f^j_{mm'}$.

In conclusion, for all $\OSp(2S+2|2S)$ cases the
minisuperspace analysis gives the same spectrum (\ref{sym}) with
the same supermultiplicities and multiplicities 
corresponding to totally symmetric tensors. 

Unfortunately, in the coset sigma model, there are many interesting fields whose dimensions tend to integers as $g_\sigma\rightarrow 0$. They are not captured by the minisuperspace approximation on the supersphere, which appears thus less useful than in the WZW case. To proceed, we will consider the $g_\sigma\rightarrow 0$ limit from a slightly different point of view.

 \subsection{Perturbation theory in the $\OSp(4|2)$ sigma model}
\label{sec:pert_th}
 
It is now most useful to recover the minisuperspace result of the
previous section from a different point of
view, using standard conformal perturbation theory in the limit of
small $g_\sigma$ (for a very useful discussion of perturbation theory in the $\OO(2)$ case, see \cite{Mudry}).

To regularize the IR and UV divergences of the theory we use a square lattice of width $L$,
spacing $a$ and a total number of sites $\mathcal{N}$.
Then, one can exponentiate the term coming from the measure and
absorb it into the action to get
\begin{align}
\label{action}
A=\frac{1}{2g_\sigma^2}\int d^2\,x
\Big\{
&2(1-\eta^1\eta^2) \partial \eta^1 \partial \eta^2 +
(1-2 \eta^{1}\eta^{2})\left[ (\partial \mu)^2 + \cos^2\mu
(\partial \xi_1)^2 + \sin^2\mu (\partial \xi_2)^2
\right] \Big. \\ \notag
&+\Big.\frac{2 g_\sigma^2}{a^2} \left[2\eta^1\eta^2 -
\log(\sin \mu\cos\mu)\right] \Big\}.
\end{align}
The role of the measure term in the effective action is to cancel the
tadpole divergences of the theory.

The perturbation theory is performed correctly by rescaling the
nonzero modes of the fields
\begin{equation}\label{correct_sep}
  \eta^\alpha (x) = \bar{\eta}^\alpha +
  g_\sigma\hat{\eta}^\alpha(x),\quad \mu(x) = \bar{\mu}+ g_\sigma
  \hat{\mu}(x), \quad \xi_\alpha(x) = \bar{\xi}_\alpha+ g_\sigma\hat{\xi}_\alpha(x).
\end{equation}
We shall treat the zero modes nonperturbatively.

In the limit $g_\sigma \to 0$ all the fields  decouple, and we get an
action with a pair of symplectic fermions and three bosons
\begin{align}\notag
 A_0 =&\, \frac{1}{2} \int d^2\,x \Big\{ 2(1-\bar{\eta}^1\bar{\eta}^2)\partial
   \hat{\eta}^{1}\partial \hat{\eta}^{2} +
   (1-2\bar{\eta}^1\bar{\eta}^2)\big[(\partial \hat{\mu})^2 +
   \cos^2 \bar{\mu} (\partial \hat{\xi}_1)^2+ \sin^2\bar{\mu}
   (\partial \hat{\xi}_2 )^2\big] \Big\}\\
   & +   2\mathcal{N}\breta_1\breta_2- \mathcal{N}\log(\sin \brmu\cos\brmu)
\label{free_act}
\end{align}
coupled to the zero modes $\bar{\eta}^1,\bar{\eta}^2$ and $\bar{\mu}$.
In the following we separate the zero modes contribution to the path
integral measure
$[d\eta^1d\eta^2d\mu d\xi_1d\xi_2] =
g_\sigma [d\hat{\eta}^1d\hat{\eta}^2d\hat{\mu}d\hat{\xi}_1d\hat{\xi}_2]
d\bar{\eta}^1d\bar{\eta}^2d\bar{\mu}d\bar{\xi}_1d\bar{\xi}_2$.
Note that one can rescale the dynamical fields
\begin{equation*}
  \tilde{\eta}^\alpha =
  \bigg(1-\frac{1}{2}\breta^1\breta^2\bigg)\heta^\alpha,\quad
  \tilde{\mu} = (1-\breta^1\breta^2) \hmu,\quad \tilde{\xi}_1 =
  \cos\brmu (1-\breta^1\breta^2) \hxi_1,\quad \tilde{\xi}_2
  =\sin\brmu (1-\breta^1\breta^2) \hxi_2
\end{equation*}
to eliminate all but one of the terms proportional to $\mathcal{N}$ in $A_0$
coming from the measure.
This is because leaving the zero modes nonintegrated is equivalent, in
the lattice regularization picture, to fixing
the fields in one site of the lattice.

Let us call \emph{partial} the correlation functions computed
without integrating the zero modes.
For instance the partial propagators are
\begin{align}\label{reg_pr}
\left< \heta^1(x) \heta^2(y) \right>_* & = ( 1 + \breta^1 \breta^2)
G(x,y)\\ \notag
\left< \hmu(x) \hmu(y) \right>_* & = - (1 + 2\breta^1\breta^2) G(x,y)\\ \notag
\left< \hxi_1(x) \hxi_1(y) \right>_* & = -\frac{1+2\breta^1\breta^2} {\cos^2 \brmu } G(x,y)\\ \notag
\left< \hxi_2(x) \hxi_2(y) \right>_* & = -\frac{1+2\breta^1\breta^2}{\sin^2 \brmu} G(x,y),
\end{align}
where 
\begin{equation}
  \label{f_p}
  G(x,y) =  - \frac{1}{V}\sum_{k\neq 0}  \frac{e^{i k(x-y)}}{4 - 
    2 \cos k_1 - 2\cos k_2} \approx  - \frac{1}{V}\sum_{k\neq 0}
  \frac{e^{i k(x-y)}}{k^2} \approx \frac{1}{2 \pi}\bigg(\log \frac{\pi
    |x-y|}{L}+\gamma\bigg).
\end{equation}
Here the sum is over all the quantized modes $k\neq 0$ in a box of
volume $V = L^2$.
We have also used the lattice regularization for
\begin{equation*}
  \square G(x,y) \underset{x\to y}{=} - \frac{1}{V}\sum_{k\neq 0} 1 =
  \frac{1}{V} - \frac{1}{a^2}
\end{equation*}
resulting from eq.~\eqref{f_p}, where $\square = - \partial_\mu \partial_\mu$.

Rescaling the fields as mentioned above one gets the partial partition function
\begin{equation}
\label{reg_pf}
  Z_*(\breta^1,\breta^2,\brmu) \propto g_\sigma (1-
  2\breta^1\breta^2) \sin \brmu \cos\brmu \,\protect{\det}'^{-1/2}  \square
\end{equation}
up to an arbitrary factor
coming form the normalization of the path integral measure
$[d\heta^1 d \heta^2 d\hmu d\hxi_1d\hxi_2]$.
Here $\det'\square$ is the regularized determinant of the Laplacian with periodic
boundary conditions in both directions.
Note the $\OSp(4|2)$ invariant integration measure on $S^{3|2}$
appearing in eq.~\eqref{reg_pf}.
One can choose the arbitrary constant in eq.~\eqref{reg_pf} so that
the full partition function
\begin{equation}
\label{part_func}
  Z_0 \propto g_\sigma |S^{3|2}|  \protect{\det}'^{-1/2}  \square
\end{equation}
is equal to the partition function of a single compactified boson
in the limit $g_\sigma\to 0$.
Here  $|S^{3|2}| = 4 \pi^2$ is the volume of the supersphere
$S^{3|2}$.
The correlation functions are then computed perturbatively according to the formula
\begin{equation}
  \label{mean}
  \left< O \right> = \frac{\int d \breta^1 d \breta^2 d \brmu d
    \brxi_1  d \brxi_2
    (1- 2 \breta^1\breta^2)\sin \brmu\cos\brmu \left<O
      e^{-A_{\text{int}}} \right>_*}{\int d \breta^1 d \breta^2\brmu
  d \brxi_1 d \brxi_2
    (1- 2 \breta^1\breta^2)\sin \brmu\cos\brmu \left<
      e^{-A_{\text{int}}} \right>_*}
\end{equation}
by developping in powers of $g_\sigma$ the term $e^{-A_\text{int}}$.

Let us see how these conventions work on the  example of the full two point
correlation function $\left<\eta^1(x)\eta^2(y)\right>_0$ in the free
field theory  with the action $A_0$.
The eqs.~(\ref{reg_pr}, \ref{reg_pf},\ref{part_func}) give
\begin{equation*}
  \left<\breta^1\breta^2\right>_0 = \frac{\int d \breta^1 d
    \breta^2 (1- 2 \breta^1 \breta^2) \breta^1 \breta^2}{\int d
      \breta^1 d\breta^2 (1- 2 \breta^1 \breta^2) } = -\frac{1}{2}
\end{equation*}
and
\begin{equation*}
  \left<\heta^1(x)\heta^2(y)\right>_0 = \frac{\int d \breta^1 d 
    \breta^2 (1- 2 \breta^1 \breta^2) (1+\breta^1 \breta^2)G(x,y)}{\int d
      \breta^1 d\breta^2 (1- 2 \breta^1 \breta^2) } = \frac{1}{2}G(x,y).
\end{equation*} 
Therefore
\begin{equation}
\label{etet0}
\left<\eta^1(x)\eta^2(y)\right>_0 = \left<\breta^1\breta^2\right>_0 +
g_\sigma^2 \left< \heta^1(x) \heta^2(y)\right>_0 = -\frac{1}{2}\Big(1-
g_\sigma^2 G(x,y)\Big).
\end{equation}

Higher order partial correlation functions can be computed according
to the general rule
\begin{align}\label{mult_ferm}
\left< \eta^1(x_1)\dots \eta^1(x_n)\eta^2(y_1)\dots
  \eta^2(y_n)\right>_*=
 &-g^{n-1} \breta^1\breta^2 \det \|\ln \eta^{i,i+1}_{j,j+1}\| \\
&+ g^n (1+ n
 \breta^1\breta^2)\sum_{\pi \in \Sym(n)} \varepsilon(\pi)G(x_1,y_{\pi(1)})\dots
 G(x_n,y_{\pi(n)})       \notag
\end{align}
where we have set $\eta^{ij}_{kl}=\frac{r_{ik}r_{jl}}{r_{il}r_{jk}}$
and $r_{ij} = |x_i-y_j|$.
The lowest order term in eq.~\eqref{mult_ferm} is typical of symplectic
fermions
\cite{Kausch,Ivaskevich} while the second is the usual Wick rule for
the dynamical components.

All correlation functions which might be of interest are between
products of fundamental fields $X^p$ in different points.
Therefore, in partial correlation functions the fields $\xi_a$, $a=
1,2$ will always appear in the form
\begin{equation}\label{multivert}
  \left< e^{i \alpha_1 \xi_a(x_1)} \dots e^{i\alpha_ n
      \xi_a(x_n)}\right>_* = e^{i\beta \brxi_a} \exp\Big(
  \frac{g_\sigma^2}{2\cos^2\brmu} (1+
  2\breta^1\breta^2)\sum_{k,l}\alpha_k\alpha_l G(x_k,x_l)\Big)
\end{equation}
where all $\alpha$'s are integers and by $G(x_k,x_k)$ we mean $G(x_k,x_k+a)$.
The integration along the zero mode $\brxi_a$ imposes the classical
``zero charge'' constraint
$\beta = \sum \alpha_k = 0$ for the nonvanishing full correlation function.
An immediate consequence of the zero charge condition is the
dependence of eq.~\eqref{multivert} only on the regularized propagator
$G_{\text{reg}}(x,y) = G(x,y) - G(0,0)=\frac{1}{2\pi}\log
\frac{|x-y|}{a}$
independent of the IR cut-off $L$.
Finally, although one has to perform a nontrivial integration for the
zero modes $\breta^1,\breta^2,\brmu$, the two point function of vertex
operators is as usual
\begin{equation}
  \label{vert_xi}
  \left< e^{i \alpha \xi_a(x)} e^{- i\alpha
      \xi_a(y)}\right>_0 =\Big|\frac{x-y}{a}\Big|^{-g\alpha^2}.
\end{equation}
Indeed, let $c = g_\sigma^2 (1+ 2
\breta^1\breta^2)G_\text{reg}(x,y)$.
Then, making the change of
variables $u = 1+\tan^2 \brmu$ and integrating by parts one can bring the correlator in
eq.~\eqref{vert_xi} to the form
\begin{equation*}
\frac{1}{2}  \int d\breta^1 d\breta^2 (1- 2\breta^1\breta^2) \left(
  e^{-c} - c \Gamma(0,c)\right),
\end{equation*}
where $\Gamma(0,c)$ is the partial gamma function.
In order to integrate the fermionic zero modes we develop
$\Gamma(0,c)$ around the body of $c$.
Then $c \Gamma(0,b(c))$ shall not contribute to the final result
and we are left only with exponentials.

On the other hand the field $\mu$ behaves quite differently.
The main reason is the fact that $\mu$ lives on a segment $0<b(\brmu)<\pi/2$ rather then
on a circle.
The integration of the zero mode $\brmu$ in the correlation function
of multiple vertex operators in $\mu$ will generate a factor 
\begin{equation*}
  I(\beta) = \int_{b(0)}^{b(\pi/2)} d\brmu\, \sin(2\brmu) e^{i \beta    \brmu} =
  \frac{2}{4 - \beta^2}\big(1+ e^{i\frac{\pi}{2}\beta}\big).
\end{equation*}
which is zero only for $\beta\in 4\mathbb{Z} +2,\, \beta\neq \pm 2$.
Therefore, there is no zero charge condition for the correlation
functions of vertex operators in the field $\mu$.
Moreover, the two point function of a vertex operator is also different
from eq.~\eqref{vert_xi} 
\begin{equation*}
\left< e^{i \alpha \mu(x)} e^{-i\alpha \mu(y)}\right>_0
= \bigg(1 + g \alpha^2 \log\frac{|x-y|}{a} \bigg)\Big|\frac{x-y}{a}\Big|^{-g\alpha^2},
\end{equation*}

Now, let us perturbate the free action $A_0$ by
the first order
\begin{align}\notag
  A_1 = - g_\sigma \int d^2\, x &\bigg\{ (\breta^1\heta^2+
    \heta^1\breta^2)\bigg[ \partial \heta^1 \partial \heta^2 +
      (\partial \hmu)^2 + \cos^2 \brmu (\partial \hxi_1)^2 +
      \sin^2\brmu (\partial \hxi_2)^2 - \frac{2}{a^2}\bigg]\bigg.\\
 & \bigg. +(1 -2 \breta^1\breta^2)\sin\brmu \cos\brmu \, \hmu \big[ (\partial
  \hxi_1)^2 - (\partial\hxi_2)^2\big] + \frac{2}{a^2}\cot(2\brmu) \,
  \hmu\bigg\}\label{a1}
\end{align}
and second order
\begin{align}\notag
  A_2 = - g_\sigma^2 \int d^2 x &\bigg\{ \heta^1\heta^2 \bigg[  \partial
  \heta^1 \partial \heta^2 + (\partial \hmu)^2 + \cos^2 \brmu (\partial \hxi_1)^2 + 
 \sin^2\brmu (\partial \hxi_2)^2 - \frac{2}{a^2}\bigg] \bigg. \\\notag
 & - \sin(2 \brmu) \, (\breta^1 \heta^2 + \heta^1 \breta^2) \hmu \big[ (\partial\hxi_1)^2 -
 (\partial \hxi_2)^2\big] \\\notag
 & + \frac{1}{2}(1- 2 \breta^1\breta^2) \cos(2 \brmu) \, \hmu^2 \big[
 (\partial\hxi_1)^2 -  (\partial \hxi_2)^2\big] \\
 & \bigg.- \frac{2}{a^2 \sin^2 (2 \brmu) } \, \hmu^2 \bigg\}
\label{a2}
\end{align}
interaction term in the action~\eqref{action}.

Because the way we compute correlation functions in eq.~\eqref{mean}
is quite different from the usual approach, it is no use in normal
ordering $\exp(-A_\text{int})$ in the numerator in
order to cancel the perturbative corrections in the denominator.
Therefore, we need the corrections to the
partition function $Z_0$ in order to compute perturbatively correlation functions.
There is no partial correction to $Z_0$ to the first order in
$g_\sigma$, that is $\left<A_1\right>_* = 0$.
The second order correction is
$Z_2 = Z_0 \left<\frac{1}{2}A_1^2- A_2\right>_0$.
In view of eq.~\eqref{reg_pf}, it is not hard to see that only the
first lines in eq.~\eqref{a1} and eq.~\eqref{a2} contribute to the
full correction after the integration of the zero modes
$\breta^1,\breta^2$.
Using integrals of the form~\eqref{3G_int}
\begin{equation*}
  \frac{1}{2} \left<A_1^2\right>_0 =  \left<A_2\right>_0 = \frac{5\mathcal{N}-3}{2}g_\sigma^2
  G(0,0)
\end{equation*}
and therefore $Z_2 = 0$.

We now have all the necessary ingredients to rederive perturbatively
the minisuperspace result of sec.~\ref{minisup} and go further in the
research of new primary operators and their scaling dimensions.

Let us first illustrate how to perturbatively compute the scaling
dimension of the six dimensional $\OSp(4|2)$ multiplet.
From eq.~\eqref{etet0}, we see that already in the free field theory
$A_0$ the fields $\eta^\alpha$ have the right scaling dimension
$g/4$.
We expect the correction to the order $g_\sigma^2$
\begin{equation}\label{explain}
  -g_\sigma \left< \left(\breta^1 \heta^2(y) +
      \heta^1(x)\breta^2\right) A_1\right>_0 + g_\sigma^2
  \breta^1\breta^2\left< \frac{1}{2}A_1^2- A_2\right>_0
\end{equation}
to be some constant proportional to $G(0,0)$.
This is indeed the case because the first term in eq.~\eqref{explain}
vanishes and the two contributions
\begin{align}
 \breta^2\breta^2 \left< \frac{1}{2} A_1^2\right>_* &= -
 \frac{1}{4}g_\sigma^2\breta^1\breta^2 \sin^2(2\brmu)\left(
   \frac{1}{\cos^4 \brmu}+\frac{1}{\sin^4\brmu}\right) \left(N-
   \frac{1}{2}\right)G(0,0)\\
\breta^1\breta^2\left< -A_2\right>_* &= 2 g_\sigma^2 \breta^1\breta^2
\left( -1 + (\mathcal{N}-1)\cot^2 (2\brmu) + \frac{\mathcal{N}}{\sin^2(2\brmu)}\right)G(0,0)
\end{align}
add up to give $-g_\sigma^2\breta^1\breta^2G(0,0)$ and we finally get
\begin{equation*}
  \left<\eta^1(x)\eta^2(y)\right> = -\frac{1}{2}\big( 1- g_\sigma^2
  G_\text{reg}\big) + g_\sigma^2 G_0.
\end{equation*}

Let us compute perturbatively the scaling dimension $h = g
\big(l+\frac{1}{2}\big)^2$, for the
highest weight  component $\eta^1 f^l_{ll}$
of the $\OSp(4|2)$ representation of highest weight
$(1,2l,2l)$ described in the
minisuperspace approach in sec.~\ref{minisup}.
From the property of the tensor product it is clear that
\begin{equation*}
  f^l_{\pm l,\pm l}(\mu,\xi_1,\xi_2)
  =f^{\frac{1}{2}}_{\pm \frac{1}{2},\pm \frac{1}{2}}(\mu,\xi_1,\xi_2)^{2l}
  =  e^{\pm i 2l \xi_1} \cos^{2l}\mu.
\end{equation*}
Separating the zero modes one has
\begin{align*}
  \left< f^l_{ll}[\mu,\xi_1,\xi_2](x)
    f^l_{-l,-l}[\mu,\xi_1,\xi_2](y)\right>_* = \exp{\bigg(-\frac{4l^2
    g_\sigma^2}{\cos^2\brmu} (1+2\breta^1\breta^2)G_\text{reg}
  \bigg)}&\times \\
\times \sum_{r,s=0}^{2l}\binom{2l}{r}\binom{2l}{s}(-1)^{r+s}\cos^{r+s}\brmu\sin^{4l-r-s}\brmu\left<
  \cos^r g_\sigma \hmu \sin^{2l-r}g_\sigma
  \hmu\cos^s g_\sigma\hmu'\sin^{2l-s}g_\sigma\hmu'\right>&_*,
\end{align*}
where we have introduced the notations $\hmu,\hmu'$ for
$\hmu(x),\hmu(y)$ and $G_\text{reg},G_0,G$ for $G_\text{reg}(x,y),
G(0,0),G(x,y)$.
The remaining correlator can be computed perturbatively by developping
in powers of $\hmu$
\begin{align}\label{free1}
 \left<   f^l_{ll}[\mu,\xi_1,\xi_2](x)f^l_{-l,-l}[\mu,\xi_1,\xi_2](y)\right>_* = 
\exp{\bigg(-\frac{4l^2     g_\sigma^2}{\cos^2\brmu}
(1+2\breta^1\breta^2)G_\text{reg}
  \bigg)} \times \\
\times \bigg\{ \cos^{4l}\brmu \big[1 + 2l g_\sigma^2
  (1+2\breta^1\breta^2)G_0\big] -
  g_\sigma^2(1+2\breta^2\breta^2)\cos^{4l-2}\brmu \sin^2\brmu\big[
  2l(2l-1)G_0 + 4l^2 G\big]\bigg\}\notag
\end{align}
To integrate the zero mode $\brmu$ one has to evaluate to the order
$a$ integrals of the
type
\begin{equation}\label{note}
  2\int_0^{\pi/2} d\brmu \sin\brmu \cos\brmu \sin^m\brmu
  \cos^n\brmu \exp\bigg(-\frac{a}{\cos^2\brmu}\bigg) =
  \frac{\Gamma\big(\frac{n}{2}+1\big)\Gamma\big(\frac{m}{2}+1\big)}{\Gamma\big(\frac{m+n}{2}+2\big)}
\left[ 1 - \frac{m+n+2}{n}a\right] + O(a^2)
\end{equation}
which are easily computed by making the change of variables $u =
\tan^2\brmu +1$ and then developping in Taylor series.
Putting everything together we finally get
\begin{equation*}
  \left<\eta^1(x)\eta^2(y)
    f^l_{ll}[\mu,\xi_1,\xi_2](x)f^l_{-l,-l}[\mu,\xi_1,\xi_2](y)\right>_0 = -\frac{1}{2(2l+1)}\left[ 1 - 2lg_\sigma^2 G_0 - g_\sigma^2 (2l+1)^2 G_\text{reg}\right],\quad l\geq \frac{1}{2}.
\end{equation*}
Thus, we get the required scaling dimension already in the free theory.

The interaction terms will contribute to the partial correlation
function with the term
\begin{equation*}
  \breta^1\breta^2 \left<\frac{A_1^2}{2}- A_2\right>_* \cos^{4l}\brmu
\end{equation*}
which gives after the integration of the zero modes the final result
for $l\geq\frac{1}{2}$
\begin{equation}\label{2pfferm}
\left<\eta^1(x)\eta^2(y)
    f^l_{ll}[\mu,\xi_1,\xi_2](x)f^l_{-l,-l}[\mu,\xi_1,\xi_2](y)\right>
  = -\frac{1}{2(2l+1)}\left[ 1  - g_\sigma^2 (2l+1)^2
    G_\text{reg} - g_\sigma^2 (2l+1)G_0\right].
\end{equation}

Repeating the same reasoning one can get the following result
\begin{align}\notag
  \left< \big[1 \mp 2j \eta^1(x)\eta^2(x) \big] \big[1 \mp 2j
    \eta^1(y)\eta^2(y)\big]
    f^{l\pm\frac{1}{2}}_{l\pm\frac{1}{2},l\pm\frac{1}{2}}(x)
 f^{l\pm\frac{1}{2}}_{-l\mp\frac{1}{2},-l\mp\frac{1}{2}}(y)\right> =\\
   \pm \big[1 - g_\sigma^2 (2l+1)^2 G_\text{reg} - (2l+1) g_\sigma^2 G_0\big]\label{2pfbos}
\end{align}
for the correlation functions between the highest and lowest weight
components of the remaining $\SO(4)\times \SSL(2)$ multiplets of the
$\OSp(4|2)$ representation of highest weight $(1,2l,2l)$.
When $j=1$ and the choice of the sign in eq.~\eqref{2pfbos}  is minus,
the factor 2 of $g_\sigma^2 G_0$ has to be corrected to 4 because of
the singularity in the gamma functions in eq.~\eqref{note}.

The perturbation theory in this section applies also to correlation functions between fields with derivatives. We give below two examples of such computations, which will be used later in sec.~\ref{sec:on_pert} where we conjecture the scaling dimension of the most general fields.

Let us compute first the anomalous dimension of the field
\begin{equation}\label{gen_field}
\eta^1 D_1\heta^1 \dots D_m \heta^1 \bar{D}_1\heta^1 \dots \bar{D}_n \heta^1
\end{equation}
where $D_i,\,  \bar{D}_j$ are holomorphic and antiholomorphic derivations of arbitrary order.
Although we switched to holomorphic and antiholomorphic coordinates we do not
consider the fundamental fields to be either purely holomorphic or purely
antiholomorphic. The field in eq.~\eqref{gen_field} is clearly a highest weight
state for a $\OSp(4|2)$ irrep of highest weight $\lambda = 1^{m+n+1}$ appearing
in the fusion of the (super)antisymmetric tensor of shape $\lambda$.

In the free theory, the partial two point correlation function for the field in eq.~\eqref{gen_field} is
\begin{equation}\label{free_contr}
\left< \eta^1 D_1\heta^1\dots \bar{D}_n \heta^1 \eta^2~'D_1\heta^2~'\dots\bar{D}_n \heta^2~'\right>_* = (-1)^{n+m}\bigg(\breta^1\breta^2 + g_\sigma^2 \big[1+(n+m+1)\breta^2\breta^2\big] G\bigg) \Gamma_0 
\end{equation} 
where $ \Gamma_0:= \left< D_1 \tilde{\eta}^1 \dots \bar{D}_n\tilde{\eta}^1 D_1\tilde{\eta}^2~' \dots \bar{D}_n\tilde{\eta}^2~'\right>$.
With the help of integrals of the type
\begin{equation}\label{3G_int}
\int_x \partial_\mu G(x_1,x)\partial_\mu G(x_2,x) G(x_3,x) = \frac{1}{2}\big( G_{13}G_{23}- G_{12} G_{23} - G_{12}G_{13}\big) +\frac{1}{32 \pi^3}
\end{equation}
it is possible to show that the correction
to the 2-point function
\begin{equation*}
-g_\sigma^2 \breta^1\breta^2\int_x \left< D_1\heta^1 \dots \bar{D}_n \heta^1 D_1\heta^2~' \dots \bar{D}_n \heta^2~' \big(\heta^1\heta^2(x)+\heta^1(x)\heta^2~'\big):\partial_\mu \heta^1(x)\partial_\mu \heta^2(x):\right>_*
\end{equation*}
coming from the perturbation $-A_1$ and susceptible to generate terms proportional to $G$, in fact, does not.
Obviously neither does $\tfrac{A_1^2}{2}$. On the other hand, the perturbation $-A_2$ induces a correction
\begin{equation}
g_\sigma^2 (-1)^{n+m}\breta^1\breta^2 \int_x \left< D_1\heta^1 \dots \bar{D}_n \heta^1 D_1\heta^2~'\dots\bar{D}_n \heta^2~' : \heta^1(x)\heta^2(x)\partial_\mu \heta^1(x)\partial_\mu \heta^2(x):\right>_*
\end{equation} 
which with the help of integrals of the type
\begin{equation*}
\int_x \partial G(x_1,x) \bar{\partial}G(x_1,x)\partial
G(x_2,x)\bar{\partial}G(x_2,x) = \dfrac{1}{(4\pi)^3 r_{12}^2 }\log
\dfrac{r^2_{12}}{2a^2},\quad a \ll r_{12}
\end{equation*}
is shown to yield the only relevant contribution
\begin{equation}\label{inter_contr}
2n m g_\sigma^2 G(-1)^{n+m}\breta^1\breta^2 \Gamma_0.
\end{equation}
It is very important to notice that this contribution exists only if we consider both holomorphic and antiholomorphic derivatives in eq.~\eqref{gen_field}.
Adding up eqs.~(\ref{free_contr},\ref{inter_contr}) we get an anomalous dimension 
\begin{equation}\label{2_example}
-\frac{g}{4}\big(2nm+n+m-1\big)
\end{equation}
for the field in eq.~\eqref{gen_field}.
It is somewhat disturbing to observe that this expression is not the Casimir of the $\OSp(4|2)$ antisymmetric representation $\lambda= 1^{n+m+1}$ except $m=n=0$ and $m=0,\, n=1$ or $m=1,\,n=0$.

It is not difficult to generalize the above calculus to find the anomalous dimension of the more general field
\begin{equation}\label{second_field}
\eta^1 D_1\heta^1 \dots D_m \heta^1 \bar{D}_1\heta^1 \dots \bar{D}_n \heta^1 P_l(\cos 2\mu),\quad l\in \mathbb{N}
\end{equation}
which is a components of the $\OSp(4|2)$ irrep $\lambda = (2l+1)1^{m+n}$ appearing in the fusion of the tensor of shape $\lambda$.
In order to eliminate the problem of computing complicate integrals with factors $\exp\tfrac{-a}{\cos^2\brmu}$, generated by the fields $\xi_1,\xi_2$, we have chosen the component $f^l_{00}[\cos\mu,\xi_1,\xi_2] = P_l(\cos 2\mu)$, where $P_l(\cos 2\mu)$ are the Legendre polynomials. The latter obey the composition formula
\begin{equation*}
 P_l[\cos(\theta_1+\theta_2)] = \sum_{k=-l}^l P_l^k(\cos\theta_1)P_l^{-k}(\cos\theta_2),
\end{equation*}
with associate Legendre functions $P_l^k(\cos \theta)$, see \cite{Vilenkin}. It can be used to get
\begin{equation*}
P_l(\cos 2\mu) = P_l(\cos2\brmu) + 2 g_\sigma P^1_l(\cos2\brmu) \hmu +\mathcal{O}(g_\sigma^2\hmu^2).
\end{equation*}
useful in computing the partial 2-point function for the field in eq.~\eqref{second_field} in the free theory
\begin{equation}\label{gen_2_free}
(-1)^{n+m}\bigg(\breta^1\breta^2 P_l(\cos2\brmu)^2 + g_\sigma^2 \big[1+(n+m+1)\breta^2\breta^2\big] P_l(\cos2\brmu)^2G - 4 g_\sigma^2 \breta^1\breta^2 P^1_l(\cos2\brmu)^2 G \bigg) \Gamma_0.
\end{equation}
With the help of the calculations leading to eq.~\eqref{inter_contr}, we easily get
\begin{equation}\label{gen_2_contr}
2n m g_\sigma^2 G(-1)^{n+m}\breta^1\breta^2 P_l(\cos2\brmu)^2\Gamma_0
\end{equation} 
for the perturbative correction containing terms proportional to $G$.
Finally, adding up eq.~(\ref{gen_2_free},\ref{gen_2_contr}) we get, after the integration of the zero modes, the anomalous dimension
\begin{equation}\label{3_example}
\frac{g}{4}\big[4 l (l+1) +1 -2nm -n -m\big].
\end{equation}
for the field in eq.~\eqref{second_field}. Notice that for $m=n=0$ eq.~\eqref{3_example} gives, as required, the scaling dimension for symmetric tensors.


\subsection{The structure of the theory as $g_\sigma^2\rightarrow 0$}
\label{sec:clas_dim} 

The content of the theory as $g_\sigma^2\rightarrow0$  can easily be found:
a similar discussion was carried out years ago in the context of sigma
models in dimension $2<d<4$ by Lang and R\"uhl in particular
\cite{Lang}. It is most convenient for this to think of a lattice
regularization of the sigma model. The basic field $\phi^i(x)$ is in the vector
representation and has dimension zero in the limit $g_\sigma^2\rightarrow 0$. 
Composite fields  are obtained by inserting basic fields  at
neighbouring points on the lattice and sending the cut-off to zero,
generating in this way combinations of derivatives contracted into
various ways.  If one wishes to avoid derivatives (and thus obtain
fields with vanishing weight as $g_\sigma^2\rightarrow 0$), one  can
only build totally symmetric tensors:  this is the content of the
minisuperspace result. Any kind of antisymmetrization requires, to
obtain a non vanishing field, to take a derivative and  gives rise to
a conformal weight of the form $\mathbb{N}+O(g_\sigma^2)$, and thus
an integer as $g_\sigma^2\rightarrow 0$.  For instance, for the $1^2$
representation, one needs to consider quantities such as
$\phi^i(x)\phi^j(y)-\phi^j(x)\phi^i(y)$. Of course, as $x\rightarrow
y$, these combinations all disappear to leading order. A non zero
contribution is obtained by considering derivatives, ie
$\phi^i\partial_\mu\phi^j-\phi^j\partial_\mu\phi^i$ (in this case, a
component of the current), whose dimensions do not vanish in the limit
$g_\sigma\rightarrow 0$.
To zero order in $g_\sigma^2$, the dimensions of the other fields are
obtained by elementary algebra. For instance, \emph{lowest} dimensional
highest weight field in the totally antisymmetric representations $1^{p}$ is of the
form $\breta^1\partial\heta^1 \bar{\partial} \heta^1 \dots
\partial^{l-1}\heta^{1}\bar{\partial}^{l-1}\heta^{1}$ if $p=2l-1$ and has an
extra $\partial^{l}\heta^1$ or $\bar{\partial}^{l}\heta^1$ if $p=2l$. Thus, its
dimension is $h_{1^p} = [\tfrac{p^2}{4}]$, where $[\cdot]$ denotes the integer
part.
In general, to the Young diagram $\lambda$
we can associate a traceless tensor composed of Young symmetrized products of $\phi^i$ in
distinct points.
After fusion the components of this tensor become fields
of dimension
\begin{equation*}
h_\lambda(g_\sigma^2=0)=\sum_{i=1}^{\lambda_1}\bigg[\frac{\lambda'^2_i}{4}\bigg],
\end{equation*}
where $\lambda'_i$ is the length of column $i$ of $\lambda$.
If the Young diagram $\lambda$ represents the $\osp(4|2)$  highest weight $\Lambda = b\epsilon_1
+a_2\epsilon_2+a_3\epsilon_3$ then
\begin{equation}
h^{min}_\Lambda(g_\sigma^2=0)=\bigg[{b^2 \over 4}\bigg ]+{|a_2-a_3|\over 2}
\end{equation}
will be the lowest possible dimension of a field in the $\osp(4|2)$ irrep $\Lambda$.

Obviously, given the same Young diagram $\lambda$, one can either use the
derivatives of $\phi^i$ in order to build  $\OSp(4|2)$ tensor
fields of higher dimension or one can multiply
the previous fields with $\OSp(4|2)$ scalars,
e.g. $\partial_\mu\phi\cdot \partial_\mu\phi$.
In this way the energy momentum tensor is a $\OSp(4|2)$ scalar
field based on the Young tableau $\emptyset$, the $\OSp(4|2)$ currents - on $1^2$ etc. 

In order to enumerate all the fields in a given $\OSp(4|2)$ irreps $\lambda$,
with the above tensor technique, one has to know what are the irreducible summands
of tensor representations. We shall be able to bypass this problem in
sec.~\ref{sec:2_conj} using the generalized symmetric function $sc_\lambda$
introduced in \cite{CanduSaleuri}.

\subsection{Perturbation theory revisited}
\label{sec:on_pert}

Using tensor techniques to carefully organize the space of states of the $\OSp(4|2)$ sigma model, it is certainly possible to extend the perturbative approach of sec.~\ref{sec:pert_th} to compute the anomalous dimension, at order $g_\sigma^2$, of arbitrary scaling fields.

However, instead of doing so it is most inspiring at this stage to recall the calculations made years ago in the context of $\OO(N)$ sphere sigma models. Most of these calculations
have been done using renormalisation group techniques after the regularization of the theory in $2+\epsilon$ dimensions.
For instance, the dimensions of symmetric
tensors (which we obtained through the minisuperspace or
perturbation theory) can  be extracted from a paper of Br\'ezin,
Zinn-Justin and Le Guillou \cite{brezin}. Their calculation was
extended to the most general case of fields involving derivatives in  a seminal work by Wegner
\cite{Wegner}.
In the latter reference, the most general fields are written in the form
\begin{equation}\label{most_gen_field}
T^{\{p\}}_{\{i\}\{k\}\{m\}\{u\}, \{j\}\{l\}\{n\}\{v\}}=
t^{p_1\cdots p_r}_{i_1\cdots i_{r_+},j_1\cdots j_{r-}}\prod_{\alpha=1}^{s_0}\big( \partial^{k_\alpha} \phi \cdot \bar{\partial}^{l_\alpha} \phi \big)
\prod_{\alpha=1}^{s_+} \big( \partial^{m_{\alpha}} \phi \cdot \partial^{n_\alpha} \phi\big) \prod_{\alpha=1}^{s_-} \big(\bar{\partial}^{u_\alpha}\phi\cdot \bar{\partial}^{v_\alpha} \phi\big).
\end{equation}
Here $t^{p_1\cdots p_r}_{i_1\cdots i_{r_+},j_1\cdots j_{r-}}$ is a \emph{traceless} tensor of rank $r=r_0+r_+ +r_-$ with $r_0$ underived fields $\phi^{p_\beta}$, $r_+$ derived fields $\partial^{i_\alpha}\phi^{p_\beta}$ and $r_-$ derived fields $\bar{\partial}^{j_\alpha}\phi^{p_\beta}$. In \cite{Wegner} Wegner claims that the tensor in eq.~\eqref{most_gen_field} has a well defined scaling dimension if it is of shape $\lambda\vdash r$ with respect to the indices $\{p\}$, of shape $\mu\vdash s_0+r_+$ with respect to $\{i,k\}$  and of shape $\nu\vdash s_0+r_-$ with respect to $\{j,l\}$. Its anomalous dimension is then
\begin{equation}\label{wegner_gen_h}
\frac{g}{4}\big[(N-1)r + 2(N-2)(s_+ + s_-) + 2\xi(\lambda)- 2\xi(\mu) - 2\xi(\nu)\big],
\end{equation} 
where $\xi$ is a function defined on partitions equal to
\begin{equation*}
\xi(\lambda) = \frac{1}{2}\sum_i \lambda_i(\lambda_i - 2 i +1).
\end{equation*}
Eq.~\eqref{wegner_gen_h} applies to arbitrary partitions $\lambda,\mu,\nu$ and has a nontrivial content even when the corresponding $\OO(N)$ scaling fields in eq.~\eqref{most_gen_field} vanish identically, that is $\lambda'_1+\lambda'_2>N$.
Given that both $\OO(N)$ and $\OSp(R|2S)$ groups leave invariant a certain symmetric scalar product, which plays a similar role in their representation theory, and that this product defines the action of both sigma models it is natural to expect that Wegner's construction~\eqref{most_gen_field} of scaling fields and computation of scaling dimensions applies to $\OSp(R|2S)$ sigma models if we let $N=R-2S$ in eq.~\eqref{wegner_gen_h}.

Let now $N=2$ and consider $\lambda = (2l+1)1^{m+n}$, $\mu = 1^m$, $\nu = 1^n$, $s_+=s_-=0$. Clearly there are no $\OO(2)$ tensors characterized by such $\lambda,\mu,\nu,s_+,s_-$ if $m+n>1$ and $l>0$ or $m+n>2$ and $l=0$.
Nonetheless, the construction~\eqref{most_gen_field} yields a nonvanishing $\OSp(4|2)$ tensor for the specified values of $\lambda,\mu,\nu,s_+,s_-$ and its anomalous dimension~\eqref{3_example} is indeed in agreement with eq.~\eqref{wegner_gen_h}.
It is hard to believe that the previous example is a mere coincidence and very tempting to conjecture
that all nonvanishing $\OSp(4|2)$ tensor scaling fields are of the form~\eqref{most_gen_field} and that  their anomalous dimensions are given by Wegner's formula with $N=2$.
However, we believe that   eq.~\eqref{wegner_gen_h} cannot  \emph{always} be right,
even in the case of $\OO(N)$ models. For instance, if $N=2$,  $s_0=s_+=s_-=0$ and $t^{\dots}_{\dots}$ is a traceless symmetric tensor of rank $l>0$ then clearly this field is a descendant of the vertex operators $e^{\pm i l \varphi}$ of dimension $h=gl^2/4$, which is different from eq.~\eqref{wegner_gen_h}.
It may be that the reason why eq.~\eqref{wegner_gen_h} is not always correct originates in the fact that zero modes of transversal coordinates $\pi_\alpha,\alpha=1,\dots,N-1$ are neglected when considering a $\OO(N)$ sigma model action of the form
\begin{equation*}
A = \frac{1}{2g_\sigma^2 }\int_x \partial \pi_\alpha \left(\delta_{\alpha\beta}+
\frac{\pi_\alpha\pi_\beta}{1-\pi_\gamma \pi_\gamma}\right)\partial \pi_\beta,\quad \phi = (\pi^1,\dots,\pi^{N-1},\sigma = \sqrt{1 - \pi_\alpha \pi_\alpha})
\end{equation*}
and using the free propagator $\left< \pi_\alpha(x)\pi_\beta(y)\right>_0 = -g_\sigma^2 G(x,y)$ in perturbation theory. The deficiency of such an approach is already obvious when comparing the propagator $\left<\phi^a(x)\phi^b(y)\right>$ at the critical point either in $2+\epsilon$ dimensions and arbitrary $N$ or in two dimensions and $N=2$, to the canonical form it must have in a conformal field theory.
The correct way to proceed would be to make a separation of the zero modes in $\pi^\alpha(x)$ and then a rescaling of dynamical components as in eq.~\eqref{correct_sep}. However, it is easy to understand that this error does not affect the computation of $\OO(N)$ \emph{scalar} correlation functions and, therefore, the computation of anomalous dimensions for the scalar fields. For instance $\left< (\phi (x)\phi(y)\right> = 1 - g_\sigma^2 (N-1)G(x,y)+\dots$ in the approach above.

Thus, the result of \cite{Wegner} referring to the locality of $\OO(N)$ scalars
\begin{equation}\label{local_fields}
\big( \partial^{m} \phi \cdot \partial^n \phi\big),\qquad  \big(\bar{\partial}^{u}\phi\cdot \bar{\partial}^{v} \phi\big)
\end{equation}
when $N=2$ must be correct. In eq.~\eqref{wegner_gen_h} we see that multiplication by such factors do not change the anomalous dimension of a field.
On the other hand, $\OO(N)$ scalars mixing $\partial,\bar{\partial}$
\begin{equation}\label{unlocal_f}
\prod_\alpha \big( \partial^{k_\alpha} \phi \cdot \partial^{l_\alpha} \phi\big)
\end{equation}
might have, according to \cite{Wegner}, nonvanishing anomalous dimension when $N=2$.
The simplest of nonlocal scalar fields is of dimension $h=4-g$ and has the form
\begin{equation}\label{prob_sc}
N(\partial \phi \cdot \bar{\partial}\phi)^2 - 2(\partial \phi
\cdot \partial\phi)(\bar{\partial}\phi \cdot \bar{\partial}\phi).
\end{equation}
It vanishes identically when $N=2$, as expected, because all scalar fields in the $\OO(2)$ sigma model are local.
In fact, in the $\OO(2)$ sigma model all the descendants of a vertex operator are created by taking derivatives of it and multiplying with scalars of the type eqs.~(\ref{local_fields},\ref{unlocal_f}).
For the $\OSp(4|2)$ sigma model, the field in eq.~\eqref{prob_sc} does no longer vanish.
Therefore, the way descendants fields are created in the conformal field theory of the $\OSp(4|2)$ sigma model might be quite different. In particular the anomalous dimension of a field would not depend only on the Casimir.


Without discussing this much further, it
is time to stress that while we have been so far discussing
corrections to the bulk spectrum, what we are really interested in is
the spectrum of the boundary theory. It is likely that the boundary conditions 
corresponding to our lattice model are  Neumann boundary conditions
 in {\sl all} directions. Note that this is a different situation
  from  the case of WZW models where the WZW term prevents, in
  string theory terms, the existence of branes that fill the entire
  background \cite{AlekseevSchomerus,Volkerfermions}. Here, we have one set of branes
  that seems to fill the whole coset - more work devoted to  boundary conditions in conformal sigma models would be required to clarify this situation entirely. It is in particular not clear at this stage how to use bulk formulas for anomalous dimensions in the boundary case, apart from the case of symmetric representations: couplings of $\partial$ and
$\bar{\partial}$ derivatives might have to be considered. We will leave this approach here and rely instead on the lattice analysis.

\section{Lattice approach to boundary $\OSp(4|2)$ sigma model}

\subsection{Block structure}

\begin{figure}
 \centerline{ \includegraphics[scale=0.5]{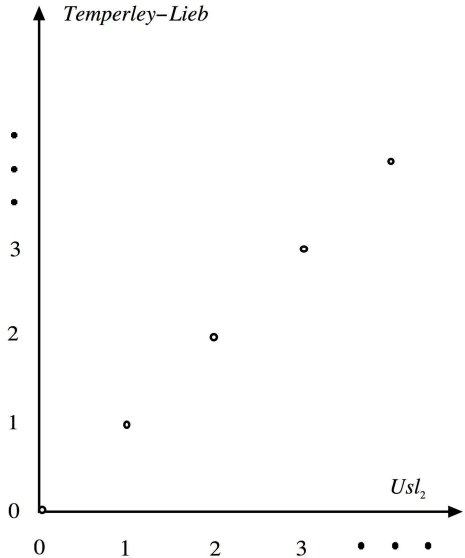}}
  \caption{In the \emph{XXX} case the decomposition of the spin chain in terms
of representations of $SU(2)$ and its commutant leads to a series of
unconnected dots. }
  \label{wzw}
\end{figure}

We remind the reader of the most important conclusion in the first paper: the
space $V^{\otimes L}$ on which the quantum hamiltonian acts can be decomposed,
for any $L$, into a series of blocks. To understand, in more physical terms the
meaning of these blocks, imagine first to study instead the \emph{XXX} spin
chain. The space $[\mathbb{C}^2]^{\otimes L}$ decomposes then as a sum of
irreducible $SU(2)$ representations times irreducible representations of the
Temperley Lieb algebra $T_L(1)$. These representations can be
indexed by a
single label, the spin $j$ which we will take integer, corresponding to $L$
even. A graphical representation of the space decomposition is given in figure
\ref{wzw}.

\begin{figure}
 \centerline{ \includegraphics[scale=0.5]{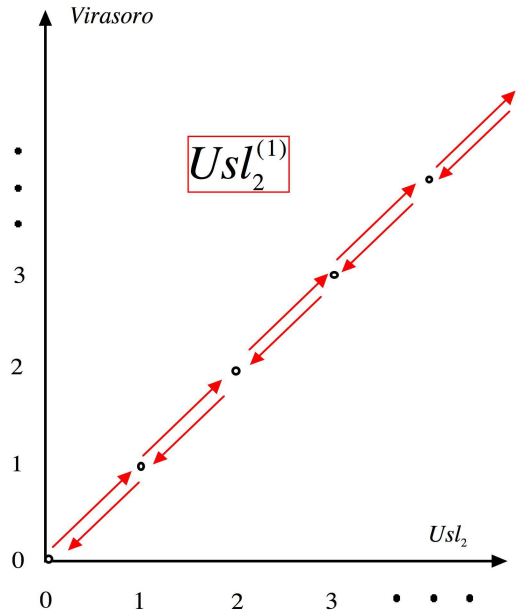}}
  \caption{In the \emph{XXX} case the dots from figure \ref{wzw} get connected
through action of the KM algebra.}
  \label{wzw1}
\end{figure}

 In the scaling limit, each representation of $T_L(q)$ is argued in
\cite{ReadSaleur07} to give rise to an irreducible representation of the
Virasoro algebra. Though this statement is not proved at the level of the full
action of the algebras, it is well established at least for the trace of
$q^{L_0}$. The trace over the $T_L(q)$ irreps $D_L(2j)$ reads thus
$$
\hbox{Tr}_{D_L(2j)}q^{L_0-c/24}= {q^{j^2}-q^{(j+1)^2}\over \eta(\tau)}
$$
Meanwhile, in the continuum limit, modes of the current algebra connect the
different $D_L(2j)$ representations, giving the Kac Moody character at level
one:
$$
\sum_{j=0}^\infty (2j+1)\hbox{Tr}_{D_L(2j)}q^{L_0-c/24}=\sum_{n=-\infty}^\infty
{q^{n^2}\over \eta(\tau)}
$$
This is illustrated as well on figure \ref{wzw1}.

The two important things to stress here are that the spin chain decomposes as a direct sum of irreducibles of $SU(2)$ and its commutant, and that in the continuum limit a Kac Moody symmetry arises.

Now for models such as ours, no Kac Moody symmetry is expected. The
Noether theorem holds, currents are conserved (at least in the bulk),
and must give rise to a multiplet in the adjoint with conformal
weights $(h,\bar{h})=(1,0)$ and another one with $(0,1)$. But the
detailed OPE's of the currents cannot obey the usual current algebra
relations, as this would imply the presence of a Kac Moody symmetry in
the spectrum, which can be excluded here  (at least for general values
of $g_\sigma^2$) from a detailed study of the degeneracies in the
spectrum and comparison with predictions based on Kac Moody
symmetry. Presumably, the OPEs of the currents are plagued with
logarithms, though few examples of such OPEs are in fact known. 

\begin{figure}
 \centerline{ \includegraphics[scale=0.5]{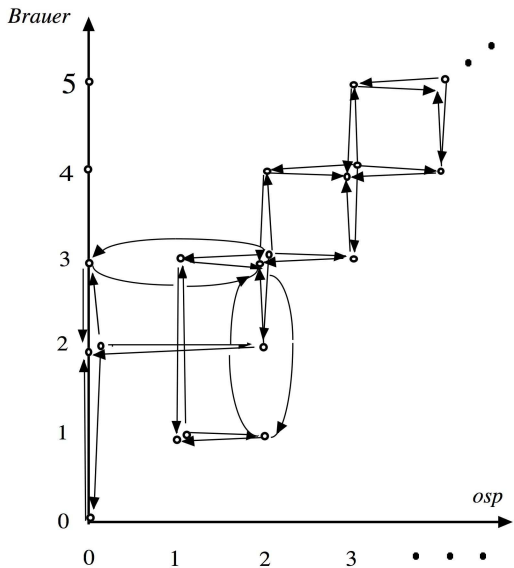}}
  \caption{In the  $\OSp$ lattice model case, representations within a block are all connected through combined action of Brauer and $\OSp$}
  \label{coset}
\end{figure}

In any case, the absence of the KM symmetry is a major inconvenient in
analyzing the $\OSp(4|2)$ supercoset sigma model. On the other hand,
the source of this complication might turn out to be our salvation as
well. Indeed, the spin chain in our case decomposes in a much more
complicated way than for the \emph{XXX} case. Representations come into
blocks, that is there are large (infinite in the scaling limit)
structures made of indecomposable representations of $\OSp(4|2)$  and
of Brauer, which are intertwined by the action of the
two algebras. The situation is similar to the ones discussed in 
\cite{ReadSaleur07} for the theories
at $c=-2$ and $c=0$ there: following  equations (4.36) and (4.37) in our first paper, blocks associated with atypical representations have the shape  shown in figure  \ref{coset}.

This shape has the nice feature (indicative of an underlying cellular
structure) that it is conserved when the length of the chain is
increased: only more nodes are added in the northeast direction, until
an infinite ladder is obtained in the scaling limit. To compare
further with \cite{ReadSaleur07}: in the case $c=-2$ they found only
one block, while in the case $c=0$ they found one block, plus an
infinity of irreducible representations not connected to the block. In
the present case, for every value of $k$, we have a block similar to
the one in \cite{ReadSaleur07}. On top of these blocks, we have many
typical representations which, like in \cite{ReadSaleur07} (see figure
8 there) are isolated, and would be represented by single dots. 

Now the crucial point is that chains such as the one we are interested
in have the same algebraic structure for arbitrary choices of the
couplings (including the running coupling constant $w$), provided they can be expressed in terms of the lattice
algebra (here the Brauer algebra). Moreover, the objects $E,\,P$ being
local must correspond in the scaling limit to local operators - and a particular combination thereof 
to the stress energy
tensor, which is clearly $\OSp(4|2)$ invariant.
Hence one
expects in the continuum limit that representations of the lattice
algebra become (for a detailed discussion of the idea see \cite{ReadSaleur07-1})
representations of an extended chiral algebra (recall that a (``fully extended'') chiral algebra in CFT is a maximal algebra of integer\footnote{
We pause here to  recall  that, like the ordinary Virasoro algebra appears in
correspondence with the Temperley Lieb algebra, ``fractional supersymmetric''
Virasoro algebras (containing generators with spin $1/k$) appear in
correspondence with representations of  the Birman Wenzl and related algebras in
integrable models based on $\ssl(2)$ spin $k=2s$ systems. In the present case
however, the identification of the $g_\sigma^2=0$ limit guarantees that only
fields with integer dimensions can appear. }
-conformal-spin holomorphic fields that have abelian monodromy and fusion rules)
 commuting with the global $\OSp(4|2)$ \emph{group} symmetry and containing the Virasoro algebra as a subalgebra. We will refer to this algebra as $\hbox{Vir}_B$, and discover some of its features as we go along.

This leads us  to the following two conjectures.

\subsection{The two conjectures}
\label{sec:2_conj}

It was shown in \cite{CanduSaleuri} sec.~4.3 that the spin chain $V^{\otimes L}$ decomposes under the action of $\OSp(4|2)$ as
\begin{equation}
_{ \OSp(4|2)}V^{\otimes L}\simeq \bigoplus_{k}D_L(k)G_{k,0} \oplus \bigoplus_{k,l}d_L(k,l) \mathcal{P}G_{k,l}\oplus \bigoplus_{\lambda\text{ typ}} d_L(\lambda)G(\lambda),
\end{equation} 
where $k=0^*,0,1,\dots$ is a label of the block $\mathcal{B}_k$ of $\OSp(4|2)$, $l=0,1,\dots$ is a label of the $l$th greatest weight in $\mathcal{B}_k$, $D_L(k), d_L(k,l),d_L(\lambda)$ are degeneracies, $G_{k,l}$ are atypical irreducible, $G(\lambda)$ are typical irreducible and $\mathcal{P}G_{k,l}$ are projective reducible $\OSp(4|2)$ representations.

We assume that the space of states of the sigma model
decomposes under the action of the global $\OSp(4|2)$ symmetry in the
same way as the spin chain $_{\OSp(4|2)}V^{\otimes L}$ decomposes in the limit $L\to \infty$.
This means that, \emph{a priori}, multiple Virasoro primary operators organize into $\OSp(4|2)$
representations~\footnote{Scaling fields with associate $\OSp(4|2)$ highest weights certainly belong to different Virasoro representations. This is because the Virasoro algebra is, in the scaling limit, a subalgebra of the lattice algebra which is $\OSp(4|2)$ invariant.} that are either irreducibles $G_{k,0}$ or
projective covers $\mathcal{P}G_{k,l}$ and typicals.
Moreover, all states within a block $\mathcal{B}_k$
give rise, in the scaling limit, to eigenvalues of the Virasoro generator $L_0$ that
\emph{differ by integers}, since they can all be connected  through the
action of  $\hbox{Vir}_B$ and $\OSp$
arrows. Hence the conjecture
\begin{eqnarray}\label{c1}
\hbox{Conjecture 1: }\hbox{ Tr}_{{\cal B}_k} q^{L_0-c/24}=q^{h_k(g_\sigma^2)}
\sum_{n=0}^\infty D_{k,n}q^n
\end{eqnarray}
with unspecified multiplicities $D_{k,n}$.
Put loosely, two representations within the same block must have
conformal weights that differ by integers or,
equivalently, conformal weights come into towers of the form:
exponent of the base depending on $g_\sigma^2$ plus integers. 

Of course, apart from these towers, we have the typical
representations, which define blocks by themselves. One can argue by the same
argument that different Virasoro operators in the same
typical representation have highest weights that differ by integers.

The  structure of the theory in the limit $g_\sigma^2 = 0$ can be carried out
exactly as in sec.~\ref{sec:clas_dim} with the only difference that
$\partial\phi$ and $\bar{\partial}\phi$ are no longer independent on the border.
Therefore, for a tensor field $\Phi$ of shape $\lambda$ the classical dimension is
\begin{equation}
h_\Phi(g_\sigma^2=0) = \sum_i \frac{\lambda'_i(\lambda'_i-1)}{2}.\label{bord_cl_dim}
\end{equation}
In order to find out more about the scaling dimensions of the fields at the base of
blocks it is useful to notice the following.
The necessary condition
for two irreps to be in the same block $\mathcal{B}$ of $\OSp(4|2)$, requiring
the eigenvalues of the Casimir to be the same, is
also sufficient.
Based on the numerical analysis of sec.~\ref{num_anal}  and the exactedness of the small
coupling expansion for symmetric representations, it is therefore
tantalizing to suggest our second conjecture,
which is stronger than the first and of course compatible with it
\begin{eqnarray}
\hbox{Conjecture 2:  }h_{\Phi}(g_\sigma^2)
=h_{\Phi}(g_\sigma^2=0)+{g_\sigma^2\over
8\pi}C(\mathcal{B}),\quad \Phi \in \mathcal{B} \label{guess}.
\end{eqnarray}
In other words, we suggest that the anomalous dimension of \emph{any boundary} field $\Phi$ in the block $\mathcal{B}$ of $\osp(4|2)$ is exactly $g C(\mathcal{B})/4$.

The two conjectures can be put together to obtain a nice form for the partition function of the sigma model.
To fix some notations let $J^0_i,J^\pm_i$  be the generators corresponding to the
even roots $2\epsilon_i,\, i=1,2,3$ of $\osp(4|2)$ normalized so that
$[J^0_i,J^\pm_i]=\pm2J^\pm_i,\, [J^+_i,J^-_i]=J^0_i$.
Then, the generalized untwisted partition function is according to eqs.~(\ref{c1},\ref{guess}) of the form
\begin{align}\label{final_form}
 Z_g(q,u,v,w) &= \tr e^{2\pi i( u J^0_1+v J^0_2 + w J^0_3)} q^{L_0-c/24}=
\sum_k q^{g k^2/4 }b_k(q) ch_k(u,v,w)\\ 
&+\sum_{k,l} q^{g k^2/4 }
a_{k,l}(q) ch_{\mathcal{P}G_{k,l}}(u,v,w) + \sum_{\lambda\text{ typ}}
q^{g C(\lambda)/4} a_\lambda(q) ch_\lambda(u,v,w).\notag
\end{align}
Here $ch(u,v,w)$ are characters of $\OSp(4|2)$ representations and $b_k(q),\, a_{k,l}(q),\,a_\lambda(q)$ denote entire functions in $q$ we will call branching functions.
Under the change of sign $u\rightarrow u+1/2$ the characters in
eq.~\eqref{final_form} become supercharacters.
According to the discusion in sec.~(\ref{sec:per_pf},\ref{sec:tw_pf})
and the fact that the superdimension of projective and typical
representations is 0 one has
\begin{equation}\label{str_red}
Z_g(q,1/2,0,0)=Z_{NN}(q),
\end{equation}
where $Z_{NN}$ is defined in (\ref{Znn}). Let $\rho$ be the outer automorphism of $\osp(4|2)$. The generalized twisted partition function defined with an insertion of $\rho$ into the trace
\begin{equation}
Z^\text{tw}_g(q,u,v,w) = \tr e^{2\pi i( u J^0_1+v J^0_2 + w J^0_3)} q^{L_0-c/24} \rho
\end{equation}
and has the same expansion as in eq.~\eqref{final_form} except that $\OSp(4|2)$ characters are evaluated on $D(u,v,w)\rho$, where $D(u,v,w)$ is $\OSp(4|2)$ supermatrix with eigenvalues $e^{\pm 2\pi i u},e^{\pm 2\pi i (v+w)},e^{\pm2\pi i (v-w)}$.
From the discussion in sec.~4.1 of \cite{CanduSaleuri} it is clear that supercharacters of $\OSp(4|2)$ projective representations vanish on $\rho$ and therefore
\begin{equation}\label{tw_str_red}
Z^\text{tw}_g(q,1/2,0,0) = Z^\text{tw}(q).
\end{equation} 
Equalities~(\ref{str_red},\ref{tw_str_red}) yield
\begin{equation*}
b_\emptyset(q) =\sum_{n=0}^\infty (-1)^n \frac{q^{n^2}}{\eta(q)},\quad
b_{1^2}(q)=1-b_\emptyset(q) = \sum_{n=1}^\infty (-1)^{n-1}
\frac{q^{n^2}}{\eta(q)},\quad
b_k(q)=\frac{1}{\eta(q)},\quad k\geq 1.
\end{equation*}
In order to find the explicit form of branching functions $a_\mathcal{P}(q)$ and
$a_\Lambda(q)$ from a lattice point of view one would have to understand how
 irreps of the
Brauer algebra decompose, in the continuum limit, into sums of Virasoro irreps. 
Otherwise according to eq.~4.4 of \cite{CanduSaleuri}
\begin{equation}\label{origins}
 Z_g(q,u,v,w) = \sum_{\lambda} sc_\lambda(D)\chi'_\lambda(q),
\end{equation}
where $\chi'_\lambda(q)$ is the contribution of the standard representation of the Brauer algebra $\Delta_L(\lambda)$ to the partition function in the continuous limit.
If we recall that $\Delta_L(\lambda)$ was constructed by trace substraction and
Young symmetrization then $sc_\lambda(D) \chi'_\lambda(q)$ is the contribution
to the partition function of all tensors of shape $\lambda$. Taking into account
the two conjectures and the fact that tensor fields of typical shape $\lambda$
are $\OSp(4|2)$ irreps, while tensor fields of atypical shape $\lambda$ contain
only $\OSp(4|2)$ atypical representations from the same block, we expect that
\begin{equation}\label{magic_trick}
 \chi'_\lambda(q) = z_\emptyset(q)\, q^{gC(\lambda)/4}\sum_{T \text{ shape } \lambda}q^T.
\end{equation}
Here $z_\emptyset(q)$ is the trace of $q^{L_0-c/24}$ evaluated on the space of $\OSp(4|2)$ invariant states constructed from products of scalars $(\partial^n \phi\, \partial^m\phi)$, $T$ is a standard Young tableau of shape $\lambda$ with entries  $n_T(\epsilon) \in  \mathbb{N}$ in every box $\epsilon\in \lambda$ and $q^T:= \prod_{\epsilon\in \lambda }q^{n_T(\epsilon)}$. The entries of $T$ denote the order of the derivatives of fields $\phi$ from which the tensor of shape $\lambda$ was constructed.

In preparation for a discussion to come in section 5 and to illustrate (\ref{magic_trick}), let us compute finally  the partition function at the (formal) value $g=2$ to the order $q^6$. The relevance of this exercise will become clear in sec.~\ref{sec:WZW}.
To the order $q^6$ we have 
\begin{equation}\label{inv}
 z_\emptyset(q) = q^{-1/24} \big(1+q^2+q^3+2q^4+2q^5+7q^6 +\dots\big).
\end{equation} 
Indeed, there is no linear term in $z_\emptyset$ because the only possible field at level 1 vanishes $(\phi\partial \phi)\equiv0$. At level 2 and 3 there is only one linearly independent field because of the constraint $(\phi\partial^2\phi) +(\partial\phi\partial\phi)=0$ and of its derivative, etc.
The typical
\begin{equation*}
 1^3, \dots, 1^{12},\, 21,\, 21^3, \dots, 21^8,\,  2^21^2,\, 2^21^3,\, 2^21^4,\, 31,\, 31^2
\end{equation*}
and atypical  weights
\begin{equation*}
\{\emptyset,2^3\}_0,\,  \{1^2, 21^2, 31^3\}_{0^*},\,  \{1, 2^21\}_1,\,  \{2, 2^2\}_2,\, \{3\}_3 
\end{equation*}
are the only ones with $h_\lambda(g=2)\leq 6$.
It is not hard to compute using def.~\eqref{magic_trick} that
\begin{align}\notag
\chi'_{r1^s}(q) &= z_{\emptyset}(q)\,  \frac{ q^{(r^2+s)/2} }
{(1-q^s)! (1-q^{r-1})! (1-q^{r+s})}\\
\chi'_{2^2s}(q) &=  z_\emptyset(q)\, \frac{ q^{4+s/2}(1-q^{s+1})}
{(1-q^2)!(1-q^{s+3})!},\label{some_triv}
\end{align} 
where we have introduced the notation $(1-q^k)!:=(1-q)\dots(1-q^k)$.
Putting together eqs.~(\ref{origins},\ref{inv},\ref{some_triv}) and 
$\chi'_{2^3}(q) = q^6 +\dots$ we get 
\begin{align} \notag
 Z_{g=2} = q^{-1/24}\big(1 &+ 6q^\frac{1}{2}+17q+38q^\frac{3}{2}+84
q^2+172q^\frac{5}{2}+325 q^3+594 q^\frac{7}{2}+1049 q^4 + 1796
q^\frac{9}{2} \big. \\
\big. &+ 3005 q^5 +4912 q^\frac{11}{2}+7877q^6+\ldots \big).\label{agreement_eq}
\end{align}

\section{Numerical analysis}
\label{num_anal}

We have investigated the validity of our two conjectures~(\ref{c1},\ref{guess}) numerically by a
detailed study of the hamiltonian $H_\Delta$ defined in eq.~(5.1) of our first
paper.

It is useful to recall that $H_\Delta$ was defined in the diagrammatic
representation of the Brauer algebra.
In sec.~3.2 of \cite{CanduSaleuri} we have explained how to construct
the standard representations  $\Delta_L(\mu),\, \mu \vdash
L-2k, k=0,1,\dots$ of the Brauer algebra $B_L(2)$ and have explicitly given the action of $B_L(2)$ on a basis of $\Delta_L(\mu)$.
We have implemented this construction numerically, thus, reducing the
problem of finding the spectrum of $H_\Delta$ to its diagonalization in each  of the standard modules $\Delta_L(\mu)$.
Once this is done, the results have to be carefully interpreted,
because only a subset of these eigenvalues actually appear in the
representation of $H_\Delta$ provided by the $\OSp(4|2)$ spin chain.
These can be found among the eigenvalues of $H_\Delta$, restricted to the standard modules $\Delta_L(\mu)$,
which contain simple summands $B_L(\mu)$ allowed to appear on
$V^{\otimes L}$ and enumerated in sec.~4.3 of \cite{CanduSaleuri}.

According to eq.~\eqref{guess}, the anomalous dimension of two multiplets
of fields assembled in two projective representations, that are associate to each other, is the same. To test numerically this aspect of the conjecture it is useful know the decomposition of  $V^{\otimes L}$ also as an $\osp(4|2)$ module.
In order to do so one has to modify eq.~4.36 of \cite{CanduSaleuri} by
i) identifying associate representations and ii) decomposing the self associate ones.

The scaling dimensions for the $\osp(4|2)$ fields in the
symmetric representations being well known from the 6
vertex model, we concentrate on projective representations only.

So, let i) $\mathcal{P}(\lambda)\neq \mathcal{P}(\lambda^*)$ be associate
projective direct summands of
$_{\OSp(4|2)}V^{\otimes L}$ paired up with the $B_L(2)$ irreps
$B_L(\lambda)$ and $B_L(\lambda^*)$  respectively.
Here $\lambda\neq \lambda^*$ are $\OSp(4|2)$ associate highest weights
induced from the $\osp(4|2)$ highest weight $\Lambda=\tau\cdot \Lambda$ invariant under
the action of the $\osp(4|2)$ outer automorphism $\tau$.
The Young tableau notation for the highest weights of $\OSp(4|2)$ was
introduced in sec.~4.3 of \cite{CanduSaleuri}.
Thus, the $_{\osp(4|2)}V^{\otimes L}$ direct summands
$\mathcal{P}(\lambda)\neq\mathcal{P}(\lambda^*)$ are isomorphic  to
$\mathcal{P}(\Lambda)$ and the latter has to be paired up with
$B_L(\lambda)\oplus B_L(\lambda^*)$ in the decomposition of
$_{\osp(4|2)}V^{\otimes L}$.

Now let ii) $\mathcal{P}(\lambda)$ be a selfassociate projective direct
summand of $_{\OSp(4|2)}V^{\otimes L}$  paired up with the $B_L(2)$
irrep $B_L(\lambda)$. Here $\lambda=\lambda^*$ is a $\OSp(4|2)$ highest weight
induced from the distinct $\osp(4|2)$ highest weights $\Lambda,
\tau\cdot\Lambda$.
Then, under the restriction to the proper subgroup of $\OSp(4|2)$,
$\mathcal{P}(\lambda)\simeq \mathcal{P}(\Lambda)\oplus
\mathcal{P}(\tau\cdot \Lambda)$.
Therefore, both  $_{\osp(4|2)}V^{\otimes L}$ direct summands
$\mathcal{P}(\Lambda)$ and $\mathcal{P}(\tau\cdot \Lambda)$ are paired
up with $B_L(\lambda)$ in the decomposition of $_{\osp(4|2)}V^{\otimes L}$.

To test the second conjecture numerically for a multiplet
of fields assembled in a $\osp(4|2)$ representation $\mathcal{P}(\Lambda)$ we compute the lowest eigenvalues $E_{\lambda}(L),E_{\lambda^*}(L)$ of $H_\Delta$
in $\Delta_L(\lambda),\, \Delta_L(\lambda^*)$ and then we compare
\begin{equation*}
h_{\lambda,\lambda^*}(L) = \frac{L}{\pi
v_s}\bigg(E_ {\lambda,\lambda^*}(L)-E_0(L)\bigg),\quad v_s = \frac{\sin \pi g}{1-g}
\end{equation*}
to eq.~\eqref{guess} with $C(\lambda)=C(\lambda^*)=C(\Lambda)$.


The efficiency of our program is such that $H_\Delta$ could be
diagonalized only in the representation spaces $\Delta_L(\mu)$'s with
$L\leq 12$.
In this range, the quantities $h_\lambda(L)$  tend to be strongly
affected by finite size corrections.
This is obvious in the limit $g\to 0$ due to the vanishing of
$v_s$ and, perhaps, less obvious in the opposite limit $g\to 1$, where the
finite size corrections (at least to the 6 vertex scaling dimensions)
are extremely slowly converging because of their logarithmic nature,
see \cite{Batchelori}.

To extrapolate the functions $h_\lambda(L)$ at $L=\infty$ we have
used the following form for the finite size corrections
\begin{equation*}
  h_\lambda(L) = h_\lambda(\infty) +
  \begin{cases}
    \frac{a}{L^2}, & C(\lambda)=0\\
    \frac{a+b\log L}{L^2}, & C(\lambda) \neq 0
\end{cases}.
\end{equation*}
Here we have implicitly
assumed that least irrelevant operators in $H_\Delta$ have scaling dimensions $2 + \mathcal{O}(g)$.
As will be seen from the figures, for the fields with $C(\lambda) = 0$, the extrapolation ``works
better'' if we drop the logarithm.

Using various symbols, we have represented in the figures bellow the values of
$h_\lambda(L)$ for every available width $L\leq 12$ and $g=0.1,0.2,\dots,1.0$.\footnote{In the vicinity of $g=0$ some $h_\lambda(L)$
are not be visible in the  figures because there are too far from
$h_\lambda(\infty)$}
To determine $h_\lambda(\infty)$ we used a least squares
fit. These values are represented in the figures by a cross.
We also draw the conjectured dependence of
$h_\lambda(\infty)$ on $g$ by a line.

Let us start our analysis with the two simplest typical $\OSp(4|2)$ Young
diagrams, that is $21$ and $1^3$. These are non self associate diagrams
and, according to \cite{CanduSaleuri} sec.~4.3 their associate
partners are $(21)^* = 3^221$ and $(1^3)^* = 32^3$.
\begin{figure}
 \begin{center}
    \begin{tabular}{rr}
   \includegraphics[scale=0.8]{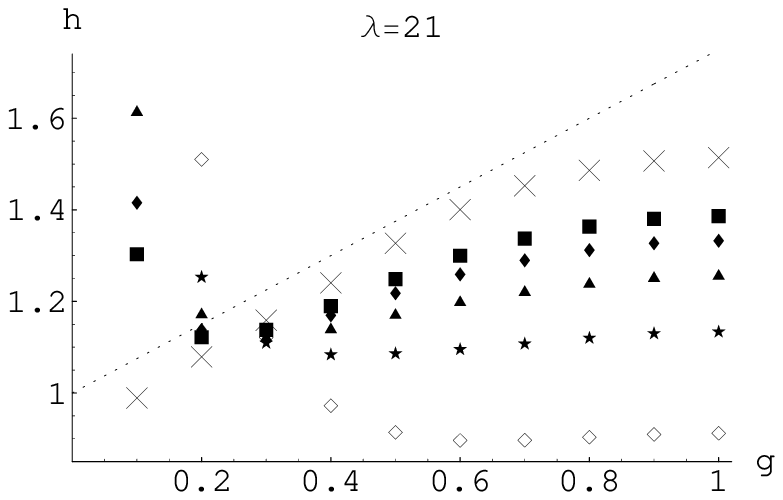} &
   \includegraphics[scale=0.8]{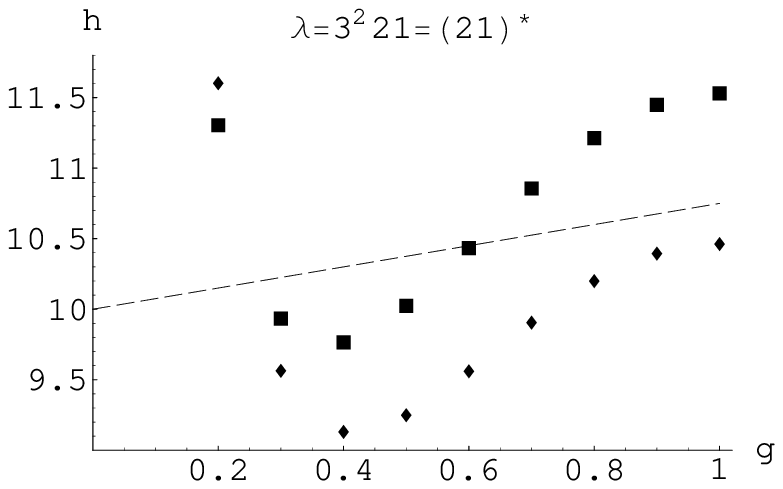} \\
   \includegraphics[scale=0.8]{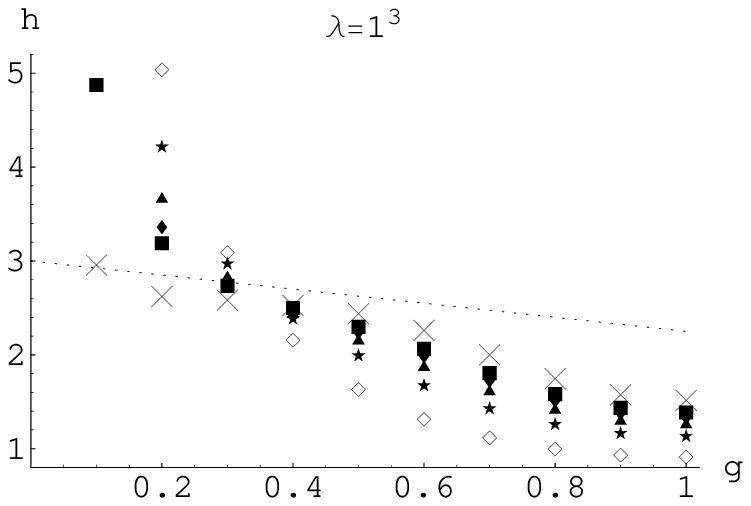} &
   \includegraphics[scale=0.8]{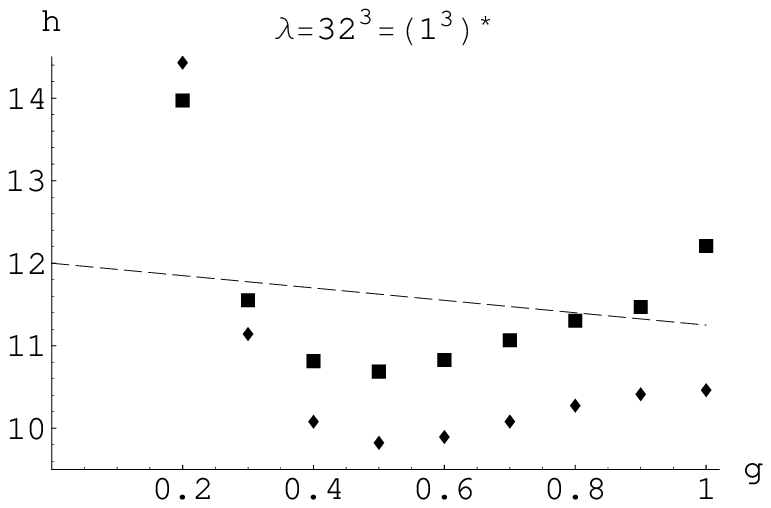}
\end{tabular}
\end{center}
\caption{Empty rhombus, stars, triangles, full rhombus and boxes represent the raw
  values of the critical exponents $h_\lambda(L)$ for $L=3,5,7,9$ and
  11 respectively; the cross
  corresponds to extrapolated values $h_\lambda(\infty)$; the
  line represents the conjectured exponents.}
  \label{fig:low_typ}
\end{figure}
The convergence of raw values $h_\lambda(L)$, the quality of
the extrapolation $h_\lambda(\infty)$ and the conjectured exponents
are exposed in fig.~\ref{fig:low_typ}.
There are no extrapolated exponents in the graph on the right in
fig.~\ref{fig:low_typ} because there are only two available widths for
the associate partitions.
Although the agreement with the second conjecture does not appear to be so impressive
at the first sight, it is worth keeping in mind that compared to Bethe
ansatz calculations, the widths we use are far smaller.
On the other hand, the classical dimension are in very good agreement with eq.~\eqref{bord_cl_dim}. As shall be seen in the following, this is the case will all the graphs we present.

\begin{figure}
  \centering
  \begin{tabular}{rr}
    \includegraphics[scale=0.8]{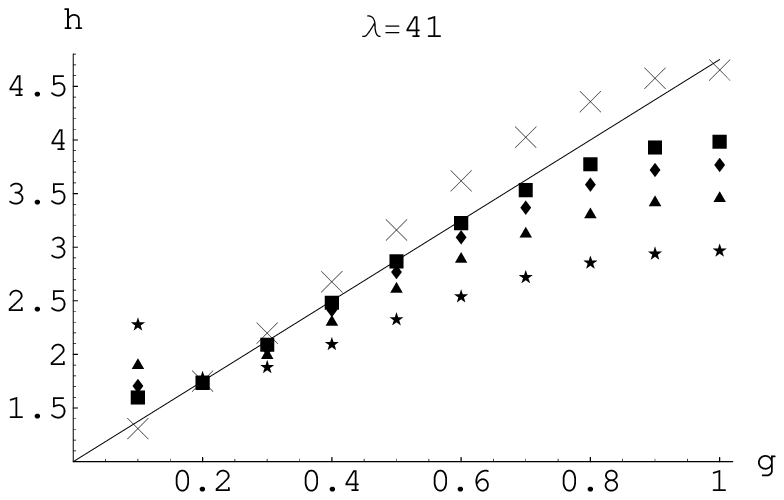} 
  & \includegraphics[scale=0.8]{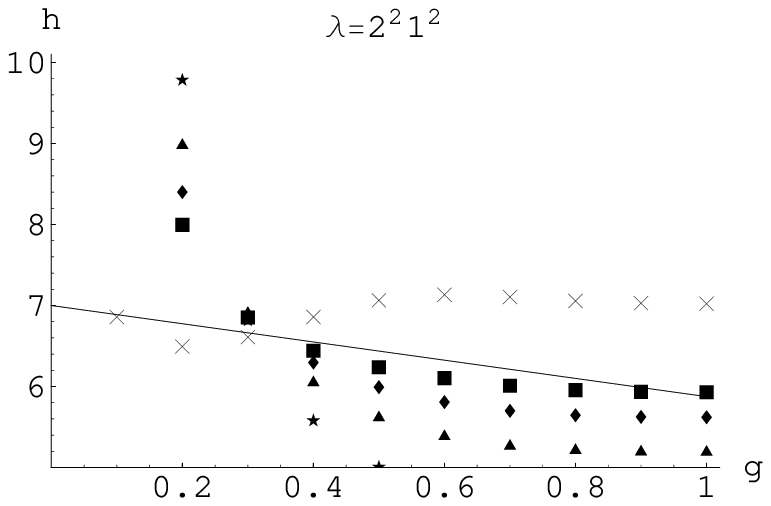}
  \end{tabular}
  \caption{The symbols we use to represent points on the left graph have the same
    meaning as in fig.~\ref{fig:low_typ}. On the right, empty rhombus,
    stars, triangles, full rhombus and boxes correspond to
    $h_\lambda(L)$ for $L=4,6,8,10$ and 12 respectively. The cross and
  the line have the same meaning as on the left.}
  \label{fig:high_typ}
\end{figure}

From our analysis of many typical Young diagrams of up to 8 boxes, which we do not present here, we have
made two interesting observations: i) $h_\lambda(L)$ converges faster for larger
partitions and ii) the lager  $C(\lambda)$ is, the faster the convergence seems to be.
We illustrate the observation i) and ii) in the graphs on the left
and, respectively, on the right in fig.~\ref{fig:high_typ}.

\begin{figure}
  \centering
  \includegraphics[scale=0.8]{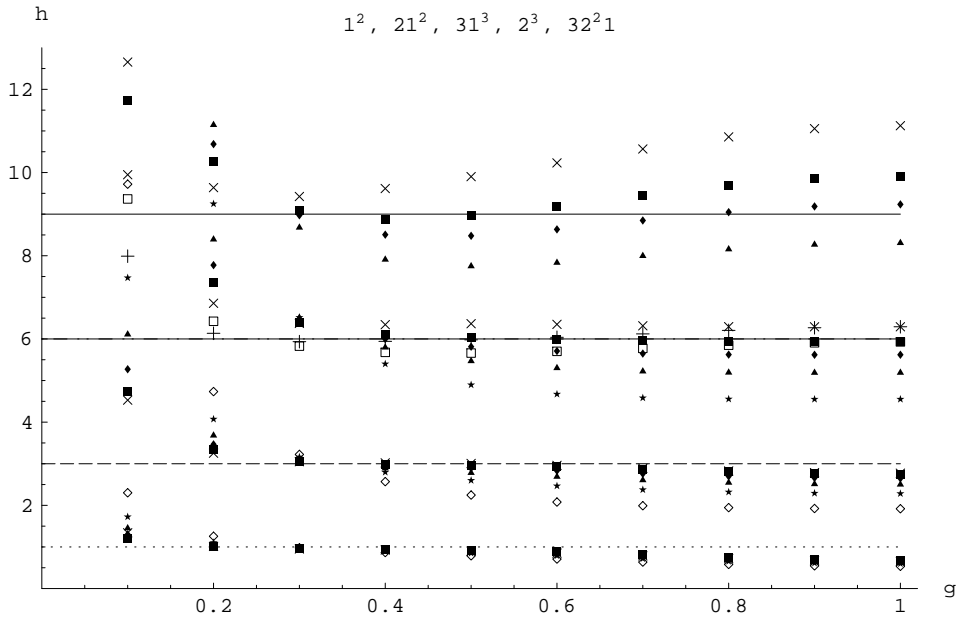}
  \caption{In order to avoid the confusing agglomeration of points
    around $h=6$ we have represented for the partition $2^3$ only
    $h_{2^3}(12)$ (empty box) and $h_{2^3}(\infty)$ (untwisted
    cross). Other symbols keep the same meaning as in
    fig.~\ref{fig:high_typ} right.}
  \label{fig:atyp_0}
\end{figure}

We turn now to the analysis of nontrivial blocks of $\OSp(4|2)$ or, equivalently, $B_L(2)$. 
First let us look at the trivial block composed of fields with
vanishing Casimir.
In order to keep the graphs in fig.~\ref{fig:atyp_0} clear, we have restricted to the
partitions $1^2,\,21^2,\,31^3$ and $(21^2)^*= 2^3,\, (31^3)^*=321^2$.
As one can see from fig.~\ref{fig:atyp_0}, the raw exponents
$h_\lambda(L)$ have a negligible dependence on $g$, as expected.
The agglomeration of points around $h=6$ might be a little bit
confusing because of the degeneracy $h_{2^3}(g_\sigma^2) = h_{31^3}(g_\sigma^2)$.

To conclude our numerical analysis we show the scaling dimension for the first four
partitions $3,\, 32,\, 3^21$ and $(3)^*=3^3$ in a block with exponents depending on $g$.
\begin{figure}
  \centering
\includegraphics[scale=0.8]{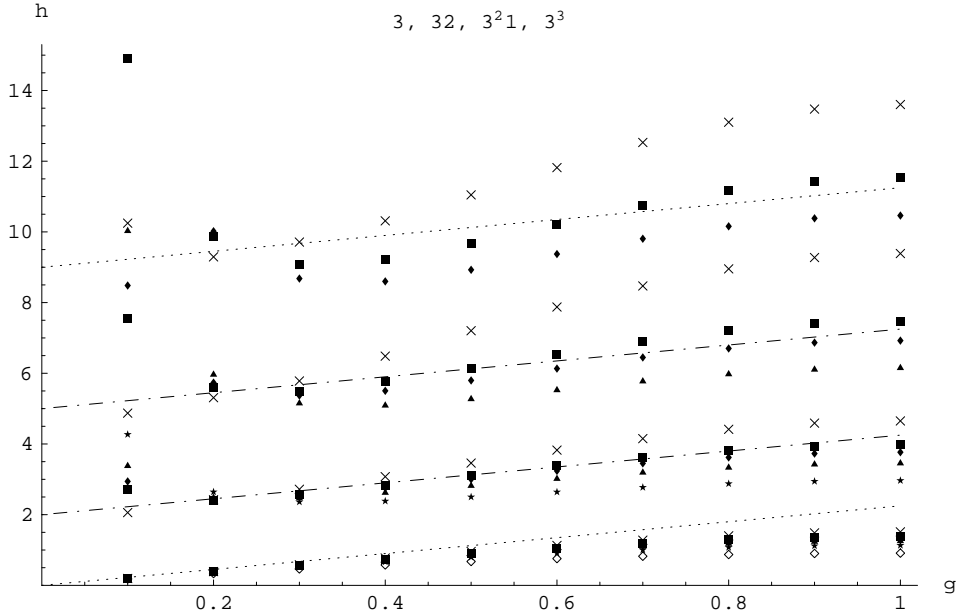}
\caption{The symbols keep the same meaning as in fig.~\ref{fig:low_typ}.} 
  \label{fig:atyp_3}
\end{figure} 

Agreement with the two conjectures is clearly quite
satisfactory in figure~\ref{fig:low_typ} to excellent in figures~\ref{fig:high_typ}, \ref{fig:atyp_0} and \ref{fig:atyp_3} for $g$ neither too close to 0 nor to 1.
It is also perfectly clear that eq.~\eqref{3_example} cannot explain the numerical results for $\lambda = 2^21^2,\, 21^2, 31^3, 2^3, 32 ^21, 3^21,3^3$.

We finally come to the point $g=1$. Things appear satisfactory at first sight. The
conjecture predicts only integer exponents for even sizes and
exponents of the form $\mathbb{N}+1/4$ for odd sizes, in agreement with the
known result. The conjecture also predicts exponents
$h_{1^p}(g=1)={p^2\over 4}$, which can be established directly at
$g=1$. Nevertheless, more subtle details do not  match. For instance,
the exponent of $\lambda = 21^2$ appears to have in fig.~\ref{fig:atyp_0} a constant value $h_{21^2}=3$ on the whole critical line, which is in agreement with the second conjecture. On the other hand, as was explained in sec.~\ref{sec:w=0} there is an exact degeneracy  $h_\lambda(L)$  at $g=1$ for all $\lambda\vdash m$.
Given that for a one raw partition $\lambda$ we know for certain that $h_m(\infty) =  m^2/4$ one would expect, in particular, $h_{21^2}(g=1)$ to be equal to 4, which is not the case: for any $g<1$,  $h_{21^2}(L)$ appears to convergence toward 3  in fig.~\ref{fig:atyp_0}. 

A closer look shows that requiring continuity of the spectrum of exponents at
$g=1$ requires giving up both conjecture 1 and conjecture 2. The
numerical results and the internal consistency of the approach suggest
rather that the point $g=1$ is in some sense singular, and that some
of its exponents (roughly, those of mixed tableaux) are discontinuous
at this point - that is, limits $w\rightarrow 0$ and $L\rightarrow \infty$ do not commute in the lattice model
(this is suggested by the bare numerical data as well). Note this is exactly what happens at $g=0$ ($w=\infty$) where the lattice model is entirely frozen and all critical exponents  - when formally defined through finite size scaling say - vanish, but the limit $g\rightarrow 0$ is well defined, and different. 

The presence of a singularity at $g=1$  may not be so surprising  if we notice that  in the lattice model, the point $w=0$  (as well as the point $w=\infty$) corresponds
to a larger symmetry ($\OSp$ enhanced to $\SU$, where the spectrum is organized according to representations 
 of the Temperley Lieb algebra, unlike the  full Brauer algebra which appears generically. In dense intersecting loop models where the fugacity of loops is $N<2$, a similar singularity also occurs at the point $w=0$. In that case however, the $w>0$ phase has less interesting features than for the case $N=2$, while the operator coupled to $w$ is relevant. Here, 
the value $g=1$ corresponds to 
some operators becoming marginal.

\section{Relation with the GN model and another look at the conjecture}
\label{sec:WZW}

The value $g=1$ plays for the 6 vertex model a role similar to the Kosterlitz Thouless point for the $\OO(2)$ model; beyond the value $w=0$, the 6 vertex model enters a massive phase, and has no longer a continuum limit of interest. We can however perfectly well continue  our formula beyond $g=1$.\footnote{
In the approach of \cite{PolchinskiMann} this continuation seems possible as well.}
In the case of the free boson, the result still describes the partition function of a well defined theory, which would be obtained as the continuum limit of a lattice model where ``the bare fugacity of vortices is turned to zero''.  Similarly, it is natural to expect that the $\OSp$ coset sigma model can be defined for any $g$ (note that for $S>0$ the supersphere has trivial homotopy).  We observe now that if we set $g=2$  or $g_\sigma^2=4\pi$ in our
conjectured formula, we obtain,
parametrizing now the Young diagram by the lengths $n_1,n_2$
of the first two rows and $b$ for the first column
\begin{eqnarray}
h(g=2)&=&{n_1^2\over 2}+{n_2^2\over 2}+{b-2\over 2},\quad b\geq 2\nonumber\\
h(g=2)&=&{n_1^2\over 2},\quad b=0,1\label{magicweights}.
\end{eqnarray}
This formula has an appealing physical interpretation. Indeed,
consider now the $\OSp(4|2)$ GN model
\begin{equation}\label{g=2act}
S=\int {d^2x\over 2\pi}\left[\sum_{i=1}^4 \psi^i_L\partial\psi^i_L+\psi^i_R\bar{\partial}\psi^i_R
+2\beta_L\partial\gamma_L+2\beta_R\bar{\partial}\gamma_R +g_{GN}
\left(\psi^i_L\psi^i_R+\beta_L\gamma_R-\gamma_L\beta_R\right)^2\right].
\end{equation}
The central charge of this model is of course $c=1$ (with a
contribution of  $2$ from the fermionic sector and $-1$ from the
bosonic one). Its beta function vanishes identically, just like in the
case of the coset sigma model.

Consider now the free point $g_{GN}=0$ where the theory reduces to four Majorana fermions and a 
$\beta\gamma$ system. The $\OO(4)$ acts on the fermions here, and the
$\SP(2)$ on the bosons, in contrast with the supersphere sigma model
where the $\OO(4)$ acts on the bosons and the $\SP(2)$ on the
(symplectic) fermions. 

In the free theory, the basic fields $\psi^i,\beta,\gamma$ all have
dimension $h={1\over 2}$, and they live in the fundamental
representation.
It is easy to organize all the fields of the
theory in terms of representations of the global $\OSp(4|2)$ symmetry by means of Young
diagrams.\footnote{We do not pretend that traceless tensors correspond
to $\OSp(4|2)$ \emph{irreps}. We only use the fact that a (traceless)
tensor of shape $\lambda$ necessarily contains the $\OSp(4|2)$ irreps of highest
weight $\lambda$ and, if $\lambda$ is atypical, it might also contains
irreps representation of highest weight $\mu<\lambda$. }
Consider a Young diagram of shape $\lambda$. To build a tensor
corresponding to $\lambda$  one has to proceed as in the case of the
sigma model, that is to super(anti)symmetrize indices~\footnote{The indices correspond
to the basis vectors $\psi^i,\beta,\gamma$ of the fundamental
representation of $\OSp(4|2)$.}  in a row(column).
In this case, however, even indices correspond to
fermionic fields $\psi^i$ (in different points), while odd indices correspond to bosonic
fields $\beta,\gamma$.
Therefore, the components of a tensor of hook shape
$\lambda=n_1n_21^{b-2},\,  n_1=(a_2+a_3)/2+1,\,  n_2 = |a_2-a_3|/2+1$ corresponding
to an $\osp(4|2)$ highest weight
$\Lambda=b\epsilon_1+a_2\epsilon_2+a_3\epsilon_3$ acquire, after the fusion, the
dimension
\begin{equation*}
 \frac{1}{2}\times \big(b+n_1+n_2-2\big)+
 \frac{n_1(n_1-1)}{2}+\frac{n_2(n_2-1)}{2}
\end{equation*}
which is exactly eq.~\eqref{magicweights}. This is, of course, 
the lowest possible dimension for a multiplet of $\osp(4|2)$ fields in
a highest weight representation $\Lambda$. All other fields with the
same symmetry $\Lambda$ have dimensions that differ by the previous
one by integers. They can be constructed in two ways: i) instead of
taking the fundamental fields themselves in order to build
tensor one can take as well their derivatives  ii) one can also
multiplying the previously considered fields
with $\osp(4|2)$ scalars, e.g. $\psi^i\partial\psi^i
+\beta\partial\gamma-\gamma\partial\beta$. 

Thus we propose that the continuation of the sigma model to $g=2$ coincides
with the GN model at $g_{GN}=0$, which is nothing but  the $\OSp(4|2)$
WZW model at level $k=-1/2$.\footnote{The fermionic fields $\psi^i$ in
  eq.~\eqref{g=2act} clearly provide a free field representation for
  $\so(4)_1\simeq \ssl(2)_1\oplus \ssl(2)_1$. The level
of the $\ssl(2)$ free field representation provided by the fields $\beta,\gamma$
can be derived from the normalization of the $\osp(4|2)$ even roots:
$(2\epsilon_1)^2= -4,\,(2\epsilon_2)^2 = (2\epsilon_3)^2 = 2$.}

To check this further, let us observe that the full organization of the fields  in the WZW 
can be obtained more explicitly  by  analyzing  the characters of the
current algebra in this theory.
One finds two representations $\{0\},\{1\}$ for the affine superalgebra based on the
trivial and fundamental representations of
$\osp(4|2)$, with characters $\chi_{\{0\},\{1\}} (\tau,u,v,w) =
\tr_{\{0\},\{1\}} q^{L_0-c/24} e^{2\pi i(u  J^0_1+ v J^0_2 + w J^0_3)}$
\begin{align}
\chi_{\{0\}}(\tau,u,v,w) &= \chi^{-1/2}_0(\tau,u)
\chi^{1}_0(\tau,v)\chi^{1}_0(\tau,w) +
\chi^{-1/2}_1(\tau,u)\chi^{1}_1(\tau,v)\chi^{1}_1(\tau,w)
\nonumber\\ \label{aff_ch}
\chi_{\{1\}}(\tau,u,v,w) &=\chi^{-1/2}_0(\tau,u)\chi^{1}_1(\tau,v)\chi^{1}_1(\tau,w) +
\chi^{-1/2}_1(\tau,u)\chi^{1}_0(\tau,v)\chi^{1}_0(\tau,w).
\end{align}
Here the $\chi^k_{0,1}$ are affine characters of $\ssl(2)_k$:
\begin{align}
\chi^{-1/2}_0(\tau,u)&={\eta(\tau)\over 2}\left[{1\over
    \theta_4(\tau,u/2)}+{1\over
    \theta_3(\tau,u/2)}\right]\nonumber\\ \notag
\chi^{-1/2}_1(\tau,u)&={\eta(\tau)\over 2}\left[{1\over
    \theta_4(\tau,u/2)}-{1\over \theta_3(\tau,u/2)}\right]\\ \notag
\chi^1_0(\tau,u) &= \frac{\theta_3(2\tau,u)}{\eta(\tau)}\\
\chi^1_1(\tau,u) &= \frac{\theta_2(2\tau,u)}{\eta(\tau)}
\end{align}
In order to compute the supercharacters one has to insert $(-1)^{J^0_1}$
into the trace, that is formally make the shift $u\to u+1/2$.
Imposing periodic boundary
conditions for both fermions and bosons yields a
supertrace partition function.
Evaluating the characters in eq.~\eqref{aff_ch} with $u=1/2,\,v=w=0$
we get, as expected from sec.~\ref{sec:per_pf}, the partition function of
the compactified boson at $g=2$.

Moreover, let $\rho$ be the outer automorphism of $\osp(4|2)$.
Then one can define twisted characters via
\begin{equation}\label{tw_ch}
\chi^\text{tw}_{\{0\},\{1\}} (\tau,u,v,w) =
\tr_{\{0\},\{1\}} q^{L_0-c/24} e^{2\pi i(u  J^0_1+ v J^0_2 + w
  J^0_3)}\rho
\end{equation}
Noticing that $\rho$ acts only on the $\so(4)$ Dynkin labels it is not
hard to prove that the $\osp(4|2)_{-1/2}$ twisted characters in
eq.~\eqref{tw_ch} are given by the same formulas~\eqref{aff_ch} except
the $\so(4)_1$ characters are now replaced by their twisted versions
\begin{equation*}
  \chi^1_0(\tau,v)\chi^1_0(\tau,w) \rightarrow
  \frac{\theta_3(4\tau,v+w)}{\eta(2\tau)},\quad \chi^1_1(\tau,v) \chi^1_1(\tau,w)\rightarrow \frac{\theta_2(4\tau,v+w)}{\eta(2\tau)}.
\end{equation*}
Here again, evaluating the $\osp(4|2)_{-1/2}$ twisted characters at $u=1/2,\,
v=w=0$ we get, as expected from sec.~\ref{sec:tw_pf}, the twisted characters of
the compactified boson $\chi^\text{tw}_{\{0\}}(\tau,1/2,0,0) =
\eta(\tau)/\eta(2\tau)$ and $\chi^\text{tw}_{\{1\}}(\tau,1/2,0,0)
\equiv 0$.

Adding up the two affine characters in eq.~\eqref{aff_ch} and developping in powers of $q$ in the point $u=v=w=0$ we get the partition function 
\begin{align} \notag
 Z_{g=2} = q^{-1/24}\big(1 &+ 6q^\frac{1}{2}+17q+38q^\frac{3}{2}+84
q^2+172q^\frac{5}{2}+325 q^3+594 q^\frac{7}{2}+1049 q^4 + 1796
q^\frac{9}{2} \big. \\
\big. &+ 3005 q^5 +4912 q^\frac{11}{2}+7877q^6+\ldots \big),
\end{align}
which is in agreement with eq.~\eqref{agreement_eq}, and weighs strongly in favor of our 
 identification of the point $g=2$.   This in turn provides another look at the conjectures. Indeed, instead of thinking of the critical line as a supersphere sigma model with running coupling constant $g_\sigma^2$, we may  think of it as a GN model with running coupling constant $g_{GN}$. The deformation of this model away from the WZW (free) point is a current current perturbation, for which methods of conformal perturbation theory can be applied somewhat more comfortably than in the sigma model case. 

We start by writing generally the current algebra as
\begin{equation}
J^\alpha(z)J^\beta(0)={k\over z^2}\eta^{\alpha\beta}+f^{\alpha\beta}_\gamma {J^\gamma(0)\over z}
\end{equation}
where the Greek labels take values in the adjoint.
The $\eta$'s and the $f$'s characterize the algebra, and one has the usual relations 
\begin{eqnarray}
\eta^{\alpha\beta}=(-)^{[\alpha][\beta]}\eta^{\beta\alpha}\nonumber\\
f^{\alpha\beta}_\gamma=-(-)^{[\alpha][\beta]}f^{\beta\alpha}_\gamma\nonumber\\
f^{\alpha\beta\gamma}=f^{\alpha\beta}_\delta \eta^{\delta\gamma}\nonumber\\
\eta_{\alpha\beta}\eta^{\beta\gamma}=\delta_\alpha^\gamma.
\end{eqnarray}
A crucial property of $\OSp(2S+2|2S)$ is that the Casimir in the adjoint vanishes:
\begin{equation}
\eta_{\mu\nu}f^{\nu\rho}_\sigma f^{\mu\sigma}_\tau=f^\rho_{\mu\sigma}f^{\mu\sigma}_\tau=
f^{\mu\sigma\rho}f_{\mu\sigma\tau}=C_{adj}\delta^\rho_\tau=0.
\end{equation}
This can also be used in the form $f_{\mu\sigma\rho}f^{\mu\sigma}_\tau=\eta_{\rho\tau} C_{adj}=0$.
Another crucial property  is that in $\OSp(2S+2|2S)$ there is  only one 
invariant rank three tensor, the structure constants $f^{\mu\nu\rho}$ (in other
words, every invariant in the 3-fold  tensor product of the adjoint is
proportional to $f$). Using these features, it seems possible to argue that,
exactly like in the case of the $\hbox{PSL}(2|2)$ sigma model
\cite{Bershadsky,Babichenko,Volkerbig,Volkerdraft}, the perturbation theory away
from the WZW point is abelian, and leads to corrections to the exponents
proportional, to all orders, to $g-2$ times the Casimir. In other words, we
expect for the 
 GN model that
\begin{equation}
h_\Phi(g_\sigma^2)=h_\Phi(g^2_\sigma=2)+{g^2_\sigma-4\pi\over 8\pi} C(\Phi)\label{hGN}
\end{equation}
which is identical with our second conjecture.
This leads us to our third
conjecture
\begin{align}\notag
\text{Conjecture 3: } & \text{The spectrum of the boundary } \OSp(4|2) \text{
sigma model coincides}\\
 & \text{with the spectrum of the boundary  Gross Neveu model (\ref{hGN}).} 
\label{c3} 
\end{align}

Finally, note that, while for   $g\leq 2$ then the identity field $(\phi\phi)=1$ has
the lowest scaling dimension $h_\emptyset = 0$ among all allowable states of the
theory, in contrast, as soon as $g>2$ the scaling dimension of tensor fields of
shape $1^p$ get arbitrarily large and negative for $p$ big enough, and the model presumably becomes unstable. Therefore, the point
$g=2$ has to be the end of the critical line of the sigma model.

\section{Conclusion}

This work can be summarized in our three conjectures (\ref{c1},\ref{guess},\ref{c3}). For each of these, we have given a more or less complete list of arguments and verifications - including numerical ones. Going beyond this, at the present stage, entails considerable difficulties, which we postpone for future work. One of the most obvious questions to tackle would be the nature of the boundary conditions inherited from the lattice discretization, the full lowest order calculation of the anomalous dimensions, and  the control of higher order calculations, which, according to our analysis should vanish for all representations, and not only the fully symmetric ones. Another very important question that requires further understanding is the nature of the algebra we have called $\hbox{Vir}_B$, which appears as the natural continuum limit of the Brauer algebra. 

An interesting output of our work and our first conjecture is that the boundary
spectrum is determined by the Casimir decomposition of the  spectrum at the
$g=0$ (or, if the third conjecture is correct, at the WZW point
$g=2$). This points to a special role played by the 
 Casimir algebra in the continuum limit, which is not obvious from the
consideration of  conserved quantities in the continuum action of the model
\cite{Bershadsky}, but, as we have explained, quite natural from the lattice
model point of view.\footnote{Recall that the Casimir algebra  is the algebra 
which commutes with the horizontal part $g$ of the affine Lie algebra $\hat{g}$
in usual WZW models. It contains the (enveloping algebra of the) Virasoro
algebra, 
but is much bigger in WZW models, since it also contains for instance the modes from the higher order singlet fields under $g$ built out of the currents \cite{Bais}. This is very similar to the observations made in \cite{Volkerdraft}. The characters of this algebra appear in general as branching functions in expansions of characters of the affine algebra $\hat{g}$ into characters  of $g$. 
They are not the same as the characters of the algebra generated by the Casimir fields, though the two are related in an involved way \cite{Bouwknegt}.} It seems natural to expect that further progress will come from investigating this question more thoroughly. 

\bigskip

\noindent{\bf Acknowledgments}: we thank J. L. Jacobsen and V. Schomerus for many useful discussions. We especially thank V. Schomerus for  sharing with us the results from 
 \cite{Volkerdraft} before publication. This work was supported by the Agence
National Pour la Recherche  under a Programme Blanc 2006 INT-AdS/CFT.

\end{document}